\documentclass[fleqn,usenatbib]{mnras}

\usepackage{newtxtext,newtxmath}

\usepackage[T1]{fontenc}

\DeclareRobustCommand{\VAN}[3]{#2}
\let\VANthebibliography\thebibliography
\def\thebibliography{\DeclareRobustCommand{\VAN}[3]{##3}\VANthebibliography}

\usepackage{graphicx}	
\usepackage{amsmath}	
\usepackage{xcolor}

\newcommand{\myRed}[1]{\textcolor{black}{#1}}

\title[Mapping the X-ray variability of GRS1915]{Mapping the X-ray variability of GRS 1915+105 with machine learning}

\author[B. J. Ricketts et al.]{
Benjamin J. Ricketts,$^{1,2,4}$ James F. Steiner, $^{2}$ Cecilia Garraffo $^{2}$, Ronald A. Remillard $^{3}$, and \newauthor Daniela Huppenkothen $^{4}$
\\
$^{1}$University of Southampton, University Rd, Highfield, Southampton, SO17 1BJ, UK\\
$^{2}$Harvard and Smithsonian Center for Astrophysics, 60 Garden St, Cambridge, MA, 02138, USA\\
$^{3}$MIT Kavli Institute for Astrophysics and Space Research, 70 Vassar St, Cambridge, MA 02139\\
$^{4}$Netherlands Institute for Space Research, Niels Bohrweg 4, 2333 CA, Leiden, Netherlands \\
}

\date{Accepted XXX. Received YYY; in original form ZZZ}

\pubyear{2022}

\begin{document}
\label{firstpage}
\pagerange{\pageref{firstpage}--\pageref{lastpage}}
\maketitle

\begin{abstract}
Black hole X-ray binary systems (BHBs) contain a close companion star accreting onto a stellar-mass black hole. A typical BHB undergoes transient outbursts during which it exhibits a sequence of long-lived spectral states, each of which is relatively stable. GRS 1915+105 is a unique BHB that exhibits an unequaled number and variety of distinct variability patterns in X-rays. Many of these patterns contain unusual behaviour not seen in other sources. These variability patterns have been sorted into different classes based on count rate and color characteristics by \myRed{previous work}. In order to remove human decision-making from the pattern-recognition process, we employ an unsupervised machine learning algorithm called an auto-encoder to learn what classifications are naturally distinct by allowing the algorithm to cluster observations. We focus on observations taken by the Rossi X-ray Timing Explorer's Proportional Counter Array. We find that the auto-encoder closely groups observations together that are classified as similar by \myRed{previous work}, but that there is reasonable grounds for defining each class as made up of components from 3 groups of distinct behaviour.
\end{abstract}



\color{black}
\section{Introduction} \label{introduction}

GRS 1915 is a black hole binary system located in the Milky way and was discovered in 1992 by WATCH on board the GRANAT mission \citep{Castro1992}. It has been found to be $8.6^{+2.0}_{-1.6}$kpc from Earth and the mass of the black hole of the system to be $12.4^{+2.0}_{-1.8} M_{\odot}$ \citep{Reid2014}. GRS 1915 is a rapidly rotating Kerr black hole with a dimensionless spin parameter of $a_{*} > 0.98$ \citep{McClintock2006} the value of which has been revisited and affirmed by \citet{Mills2021} and \citet{miller2013}. It is also one of the more famous BHXRB systems owing to its unusual decades-long outburst, its prominence as a source of relativistic jets, its complex sequence of X-ray variability, and its long orbital period \citep{Morgan1997,Belloni1997a,Belloni1997b,Muno1999,Belloni2000}. GRS 1915's dynamic and intricate behaviour has been a touchstone in exploring the link between accretion and ejection of material around a black hole.

\begin{figure}
    \centering
    \includegraphics[width=\linewidth]{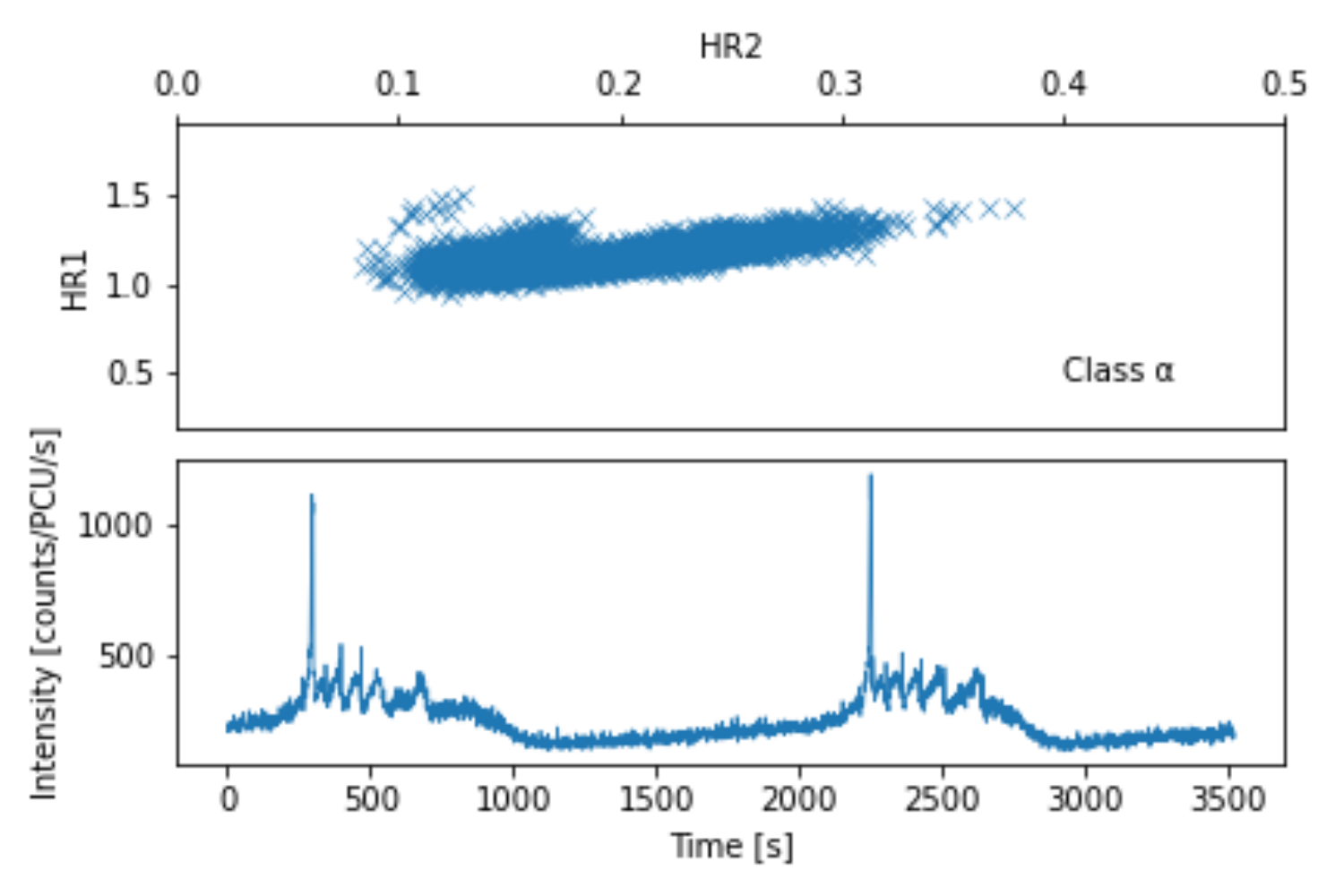}
    \caption{An example light curve and color-color diagram of the $\alpha$ class as defined by B00. The top panel is the color-color diagram. The x-axis shows points in HR2 as defined in eq. \ref{eq:HR2} and the y-axis shows points in HR1 as defined in eq. \ref{eq:HR1}. The bottom panel shows the light curve. The x-axis shows time in seconds and the y-axis shows the intensity in counts per PCU per second.}
    \label{fig:alpha_example}
\end{figure}

The large variability in the X-ray emission of GRS 1915 has been subject to considerable study. Quasi-periodic oscillations (herafter QPOs) have been observed with frequencies ranging between $10^{-3}$ and $67$Hz \citep{Morgan1997}. While the source shows exceedingly complex variability patterns, light curves and color-color diagrams contain many particular behaviours that repeat in different observations. These behaviours have been classified into 14 different classes. The first 12 were determined by \citet{Belloni2000} (hereafter B00), and two more were defined subsequently: class $\omega$ by \citet{klein2002} and class $\xi$ by \citet{Hannikaienen2005}.\footnote{the $\xi$ class is not discussed in this paper as the labeling of observations only used the first 13 classes.} An example of the one of the classes \myRed{($\alpha$)} can be seen in Figure \ref{fig:alpha_example}. The color patterns and light-curve structures used to specify each class are defined using the RXTE PCA instrument. The associated definitions would generally differ in another instrument with different energy-dependent sensitivity.

This paper focuses on analysis of observations of the black hole binary system GRS 1915+105 (hereafter GRS 1915) performed by the Proportional Counter Array (PCA) on the Rossi X-ray Timing Explorer (RXTE). These observations allow us to investigate GRS 1915's X-ray variability. Light-curve and Fourier power-density structures including QPOs are revealing about the the accretion process including for assessing potential instabilities. The behaviour of the accretion flow can change substantially during a BH outburst. Commonly, this behaviour is grouped into three canonical states: A hard state which is generally characterized by a dominant power-law component of emission thought to be produced by Compton scattering; a soft state which is associated with dominant (soft) thermal disk emission and weak power-law emission; and intermediate states during transitions between these two \citep{fender2004}. The unique variability cycles of GRS 1915 have been successfully described by B00 via transitions between three micro-states which may bear some relation to the canonical states. One of these is hard, and the other two soft. The soft micro-states are distinguished from one another by their distinctive loci and clustering properties in color-color diagrams, and which can be understood to physically correspond to different characteristic temperatures of the accretion disk associated with different accretion rates. As the source rapidly varies between these 3 states, it is crucial to understand the changes in the source to link together the source's properties to standard black hole and accretion disk models.

For a considerable period of time, GRS 1915's behaviour was unique but there has been another source identified that exhibits some of the same variability patterns as GRS 1915: IGR J17091–3624 \citep{Altamirano2011}. IGR J17091 has been found to exhibit 7 of the 14 classes observed in GRS 1915, albeit occurring at count rates between 10-50 times lower than observed in GRS 1915.

Application of machine learning techniques in astronomy has become increasingly popular in recent years, largely due to the increasingly large quantity of data that is becoming available e.g. \citep{BALL2010,fluke2020}. In the case of GRS 1915, the large volumes of observations make it a prime candidate for machine learning. Machine learning allows autonomous discovery of the most important behaviours within the data without relying on human guidance. These behaviours can be complex and difficult to quantify via a simple analytical model. Another added benefit of utilizing a machine learning model is the ability to quantify the relative importance of different behaviours in the data set for differentiating observations. When a machine learning algorithm learns how to differentiate two pieces of data, we can identify what the algorithm considers most important in the original data by looking at gradients of transformation applied by the algorithm. This may allow us to find commonalities between the classes of behaviour. Commonalities between classes of behaviour could be linked to physical phenomena. By focusing on these common behaviours, we may be able to link them together to result in the emergent and varied behaviour that we observe.

Machine learning has been applied to the case of GRS 1915 before (see \citet{huppenkothen2017,orwatkapola2022}, henceforth H17 and OK22). These authors utilise supervised learning in the form of logistic regression and random forests to quickly classify the whole RXTE data set on GRS 1915. OK22 proposed a technique using GRS 1915 as an example source; they employed a pipeline of a neural network with a long-short term memory variational auto-encoder architecture and a Gaussian mixture model to extract X-ray features for efficient classification of light curves. Both papers were able to classify (using the B00 system) light curve observations to a high level of agreement with human labeling. OK22 and H17 both sought to perform direct classification as opposed to the more continuous approach in this paper.

In this paper, we utilise a machine learning technique called an auto-encoder to gain more insight into GRS 1915's behaviour by examining GRS 1915's brightness and color-color variability. An auto-encoder can be briefly described as a neural network with a bottleneck. We compare the grouping of observations by the auto-encoder and that of a human observer. We use this framing to see where humans and the algorithm agree on important behaviours and discuss whether the current grouping of classifications is justified or can be condensed into a simpler model. Section \ref{methodology} outlines the pre-processing of observations (section \ref{data_processing}), the implementation of machine learning (section \ref{encoder}) and the use of clustering to interpret the results of the machine learning network (section \ref{clustering}). Section \ref{Results} describes the results of our method and how our network placed observations. Section \ref{discussion} further analyses the results and discusses the implications. In section \ref{conclusions}, we present our conclusions and identify potential areas for further investigation.

\section{Methodology} \label{methodology}

This section describes the preprocessing that the data underwent before being fed into the neural network as well as how the network functions.

\subsection{Data Preprocessing} \label{data_processing}
We utilised the entire data set of observations taken by RXTE's Proportional Counter Array (PCA) of GRS 1915 which amounts to a total of 2412 continuous observations. Each observation was split into 3 light curves of differing energy bands (1-s time resolution) as in B00. These three bands were: $a$: 2-5 keV, $b$: 5-13 keV, $c$: 13-60keV. From these 3 light curves, we calculated two X-ray colors:
\begin{equation} \label{eq:HR1}
    \centering
    HR_{1} = \frac{b}{a}
\end{equation}
\begin{equation}\label{eq:HR2}
    \centering
    HR_{2} = \frac{c}{a}
\end{equation}
and the total count rate ($=a+b+c$). For the background, we subtracted 4, 5 and 12 counts per proportional count unit (PCU) in the $a$, $b$ and $c$ bands respectively. This was based on the standard and updated PCA background model, released through HEASoft \citep{markwardt2009}. 

\myRed{Changes in PCA detector performance include gradual detector evolution, as well as distinct epochs in which sudden changes are caused by revised gain settings or the loss of the gas in the propane layer that sits above the xenon layer.  Calibrations for gain and detector response, for all observation times, are built into the analysis software tools by the PCA instrument team. To normalize the count rates, per second, for the chosen energy bands, we first used the channel-to-keV gain maps, for each detector epoch,  and we adjusted the channel intervals for the fast data modes (i.e., faster than 1 s) to maintain, as closely as possible, the targeted range in keV.  If the telemetry data modes provided channel boundaries that were skewed from a target interval by more than a few channels, as happened for only a handful of cases, then such observations were ignored.  The PCA background model was used to determine the background spectrum on the timescale of each continuous exposure, and the average background rates in the adopted channel ranges were subtracted from the 1-s light curves. Then, we applied a final normalization step that adjusts the throughput for each energy range and detector epoch, using contemporaneous observations of the Crab Nebula.  Within each detector-epoch, we obtained coefficients for a linear fit to the Crab count rates, integrating over energy channels in the same manner used for GRS1915, while using the Crab and Crab-background spectra per continuous exposure.  The derived coefficients were then used to normalize the light curves for GRS1915, and the final light curves are averaged over the number of PCUs active in a given observation.  The normalization uncertainty is estimated to be several percent, which is substantially less than the pre-normalization discontinuities in the raw light curves at the boundaries of detector epochs.}

Each observation has been considered by hand and, when possible, assigned to one of the first 13 classifications (i.e., omitting the $\zeta$ class). Omission of the $\zeta$ class was due to the majority of the labelling being performed prior to, or without knowledge of, the publication of \citep{Hannikaienen2005}. In cases where we were unsure of labels, we assigned it a most likely label but separated these observations from the observations which were unambiguous. 58 observations were either ambiguous in which class was best to classify them as or contained more than one class of behaviour in them (i.e. there was a transition of behaviour). These assignments are used to compare human and machine classifications.

We split each observation into segments with a length of 256 seconds. Each segment was started 8 seconds after the beginning of the previous segment such that the segments almost totally overlap, creating a running window effect. The purpose of this data augmentation is to allow the network to learn shapes regardless of their phase in the light curve, by training with similar curves at many different points in its phase. After this was performed, the data set compromised 250688 segments. By comparison, there are 8208 unique 256s segments (i.e., with no mutual overlap). It should be noted that observations can vary in length considerably and are generally much longer than the 256 second long segments that we utilise. We also found that some observations contained a short period of a drastic drop in intensity at the end of the observation. We determined this to be a result of RXTE slewing away from the target at the end of the intended observation and trimmed observations with this effect to not include this effect.

If distinguishing characteristics of one of the classes operates over periods much longer than the 256s interval, the network’s ability to identify that class is likely to be weakened. For example, the $\alpha$ class features long stretches which are relatively stable as well as other intervals exhibiting large-scale of rapid oscillations in activity. These types of drastic differences in behaviour at different times within the same class can make it hard for the network to relate them together directly. This could cause the network to separate behaviours that are directly related to one another e.g. in $\alpha$ a sharp increase in intensity is preluded by a 1800 second long slow increase in intensity. The network is unable to make the connection that a small but steady increase in intensity is an $\alpha$ class. This will lead to intrinsic differences in how the network sorts observations and \myRed{how a} human observer does.

Each observation's intensity is normalised to the range 0 to 1: where 0 is the minimum count of the observation subtracted the Poisson noise and 1 is the maximum intensity count of the observation adding the Poisson error. We find that, compared to using the (unnormalized) rate values directly, this allows the network to better reconstruct and latch onto the behaviour of low intensity light curves as well as high intensity light curves. We found that adding average intensity as a feature to the latent variables did not produce meaningfully differing results.

While it is most common to split a data set in machine learning into several different sets for the purposes of training, validation, and testing, we utilise the entire data set in training. This choice, somewhat common in unsupervised clustering, was made in an attempt to find outlying segments of light curves that may reveal a novel process or event that had been previously overlooked. As the goal of this auto-encoder is not direct classification or prediction and has not been trained with any other external information, exposing the network to "new" data for testing purposes is not necessary. 

We have also explored creating an alternative form of the data set that was balanced such that each of the classes is represented approximately equally within the data set. The balancing was achieved by adjusting the interval time that creates the ``running window'' effect. The adjustments were applied such that each class was represented by approximately between 10000 and 20000 segments. This adjustment did not make a meaningful difference in the UMAP structure (see section \ref{UMAPres}) but was used in the closeness assessment (see section \ref{closeness}) due to the prevalence of the $\chi$ class in observations. Ultimately, we determined that standard data set had lowest loss over the whole data set while still minimizing loss for each class during training. A breakdown of loss by class can be seen in Table \ref{tab:reconstruction_errs}.

\color{black}
\subsection{Auto-encoder} \label{encoder}

\begin{figure}
    \centering
    \includegraphics[width=\linewidth]{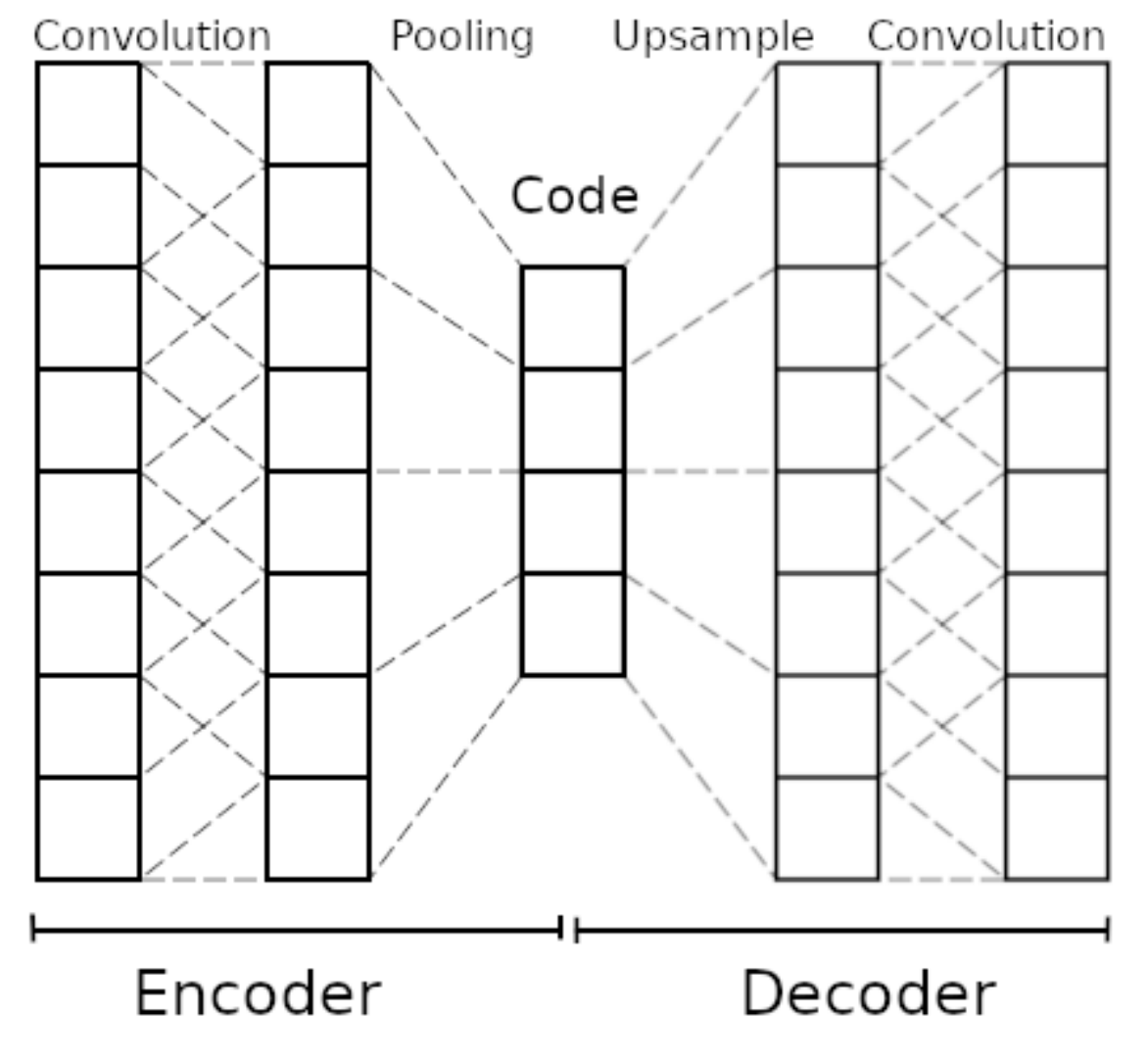}
    \caption{A cartoon of how an auto-encoder works. The input into the auto-encoder is reduced in size from its original size by the encoder to a bottleneck called a code. The code is then put through a decoder in an attempt to reconstruct the original input. The auto-encoder is trained on its ability to reconstruct the original input. When the auto-encoder is able to reconstruct the original input well, we have then produced an encoder which captures the most important behaviours from the input. We can then use this code to compare complex inputs to each other.}
    \label{fig:architecture}
\end{figure}

An auto-encoder (illustrated in Fig. \ref{fig:architecture}) is a machine learning algorithm that is able to learn robust representations of unlabeled data and can be used to dimensionally reduce inputted information down to a small latent space representation (which shall subsequently be referred to as latent variables) in a process called encoding \citep{Goodfellow2016}. To train this technique, a corresponding "decoder" is used to recreate the original input from the code produced by the encoder. It follows that, if the decoder can recreate the original input, the latent variables contains the most important information that represents the original input. This allows for the parameterization of complex behaviour into quantifiable numerical values. If the latent variables represent variability properties of each light curve, then the Euclidean distance between them is a measure of similarity between the original light curve segments.

Training an auto-encoder is performed using a loss function which quantifies the difference between the input data and the reproduction by the auto-encoder. The goal of the auto-encoder is to minimize the loss function by adjusting parameters in the layers of mathematical operations that make up the auto-encoder. When computing the loss, we are able to obtain gradients of the loss as a function of the parameters. We can then use these gradients to search the parameter space for the optimal set of parameters for minimizing the loss function, which, accordingly, is maximising the accuracy of the model. 

We utilised this technique to reduce 3 channels of 256 second time domain light curves (intensity, and two color ratios) into 48 corresponding latent variables (16 latent variables per channel of information). We first trained a network on the intensity and then retrained the resulting network on the data set for the color ratio time series, resulting in three different networks which retain structural similarities. To implement this technique, we used the Pytorch package \citep{NEURIPS2019_9015}, which is an open source machine learning framework for Python.

\subsubsection{Auto-encoder loss function}

The loss function that we used is defined as the sum of the mean square error of each output value compared to the input value weighted by uncertainty: 
\begin{equation} \label{eq:wmse}
    \centering
    \text{WMSE} = \frac{1}{N} \sum^{N}_{i=1} (\frac{x_i-\hat{x}_o}{\sigma})^{2}
\end{equation}
where \(N\) is the total number of input values, \(x\) is the input value, \(\hat{x}\) is the output value and \(\sigma\) is the uncertainty in the input value. We find that the network is able to learn long time scale behaviour well but can struggle to pick up short time scale behaviour (see the top plot of Fig. \ref{fig:reconstruct} for an example). Relatively small gains in loss obtained by modelling short time scale behaviour are difficult to encourage as often drastic changes to the model must occur. We discuss ways to combat this in further work in Section \ref{limitations}.

\subsubsection{Auto-encoder Architecture} \label{architecture}
\begin{figure}
    \centering
    \includegraphics[height=0.95\paperwidth]{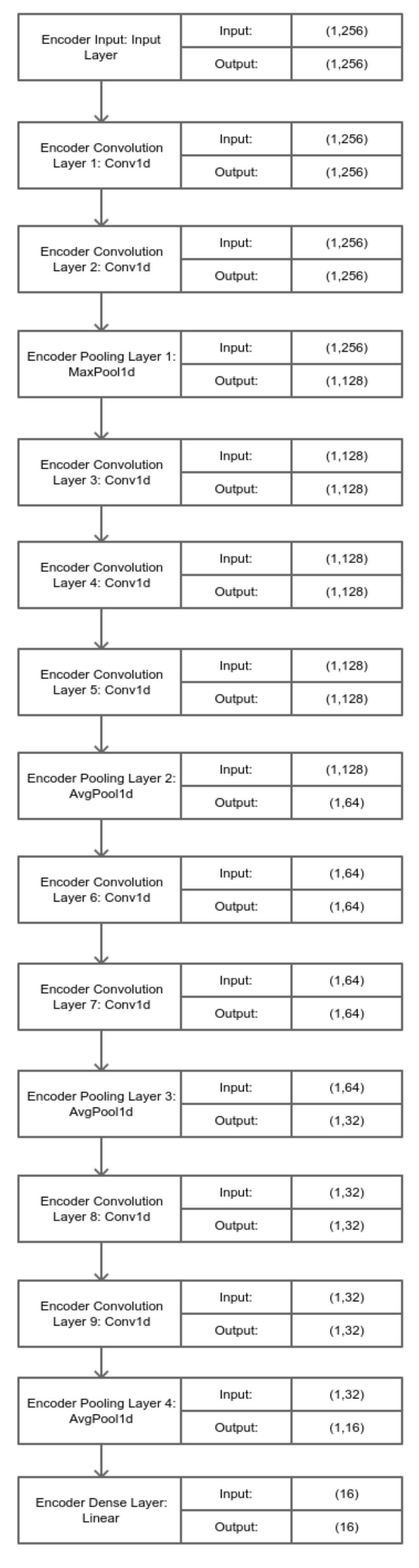}
    \caption{The architecture of the encoding portion of the auto-encoder. Left-most cell of each block indicates the label given to the corresponding instance of Pytorch layer, followed by the class of Pytorch layer. Right-most cells indicate the shape of the input and output of data of the layer. Shapes follow the convention of (channels,width).}
    \label{fig:flowchart}
\end{figure}

An auto-encoder has two major parts: an encoder and a decoder. Both of the encoder and decoder are convolutional neural networks.
Fig. \ref{fig:flowchart} shows the layers utilized in the encoder. The top cells indicate the input layer of normalized data and how data flows through the layers, with each layer applying a mathematical, non-linear operation. 

Convolution layers extract features from the input data. Using a kernel, they look at a small portion of the input data at a time and apply a series of multiplicative operations that are determined by weights to the data that appears within its kernel. They act as a filter that is translationally invariant. As an example, a convolutional layer may act as a filter looking for sharp increases in intensity in the centre of its kernel. The layer will output a high value if a high intensity peak is within the kernel and a low value if one is not observed. The first layers extract local behaviour and deeper ones extract the higher level, global behaviour.  

Pooling layers were utilised in the reduction of the size of the data. There are two different types of pooling layers: a maximum pooling layer which merely takes the maximum value within its kernel as its output, and an average pooling layer that outputs the average of the values within its kernel. These are often combined with other types of layers in what is called a block. If combined with a convolution layer like the example above, a maximum pooling layer can act similar to a delta Dirac function (the maximum value will be high if there is a peak of intensity in the kernel and thus the output of the pooling layer will be a high value) while an average pooling layer can smooth out observed behaviour (many peaks are associated with many high values so a high average but a single peak will be averaged out by its surroundings to be much smaller).

A fully connected layer is the simplest of the types of layers. It merely takes all its input variables, multiplies each one by a parameter called weight, adds a parameter called bias (these parameters are different for each input value) and adds them together to each of the output variables. 

The final layer of the encoder, before outputting the latent variables, is a fully connected layer. The fully connected layer combines the high level features that have been extracted into the latent variables that we will then use in the decoder to reproduce the original inputs.

The output of the encoder are 16 latent variables that encode the information in the light curve. These latent variables can then be used to reconstruct the light curve and compare it to the original. The number of latent variables is itself a tunable parameter. We chose 16 as we found it to be the parsimonious size still allowing effective reconstruction (e.g a low loss value) over the whole data set. We prioritized making sure that each group of observations as defined by the B00 system had comparable levels of reconstruction loss so that no one particular set of behaviour was over fitted. \myRed{Both our balanced and the original, unbalanced network perform very similarly to one another in terms of reconstruction and in UMAP projection.  This indicates that the B00 classification system hasn't been artificially imprinted onto our network in the balanced network case, and conversely that the original network's mapping is not merely driven by a select few most-common (most probable) classes. A breakdown of reconstruction loss by class can be found in Table \ref{tab:reconstruction_errs} in the appendices.}

\begin{table}
\centering
\begin{tabular}{ |c|c|c|c| }
 \hline
 Layer & Kernel size & Stride & Padding \\ 
 \hline
 Convolution 1 & 5 & 1 & 2\\  
 Convolution 2 & 5 & 1 & 2\\  
 Convolution 3 & 5 & 1 & 2\\  
 Convolution 4 & 5 & 1 & 2\\  
 Convolution 5 & 5 & 1 & 2\\  
 Convolution 6 & 3 & 1 & 1\\  
 Convolution 7 & 3 & 1 & 1\\  
 Convolution 8 & 3 & 1 & 1\\  
 Convolution 9 & 3 & 1 & 1\\  
 Max Pooling 1 & 2 & 2 & 0\\
 Average Pooling 1 & 2 & 2 & 0\\
 Average Pooling 2 & 2 & 2 & 0\\
 Average Pooling 3 & 2 & 2 & 0\\
  \hline
\end{tabular}
\caption{Table of specified parameters of each layer in the encoder. The layer name is indicated on the left, followed by the kernel size, stride of the kernel and the padding applied to each end of the output of each layer.}
\label{tab:properties}
\end{table} 

Table \ref{tab:properties} shows each of the specified parameters used in each of the layers of the encoder. Padding was used to avoid loss of information at the beginning and tail end of each light curve.

The last piece of the auto-encoder is the decoder. This has the exact reverse architecture of Fig. \ref{fig:flowchart} with two changes: Conv1d layers are replaced with ConvTranspose1d layers and both MaxPool1d and AvgPool1d are replaced with the Upsample from Pytorch. The final output layer of the decoder aims to reproduce the original data placed in the input layer. The auto-encoder is then trained to minimize the differences between the reconstructed light-curve and the original.

\subsubsection{Auto-encoder Optimiser}\label{optimiser}

\begin{figure*}
    \centering
    \includegraphics[width=0.8\textwidth,height=0.4\textheight]{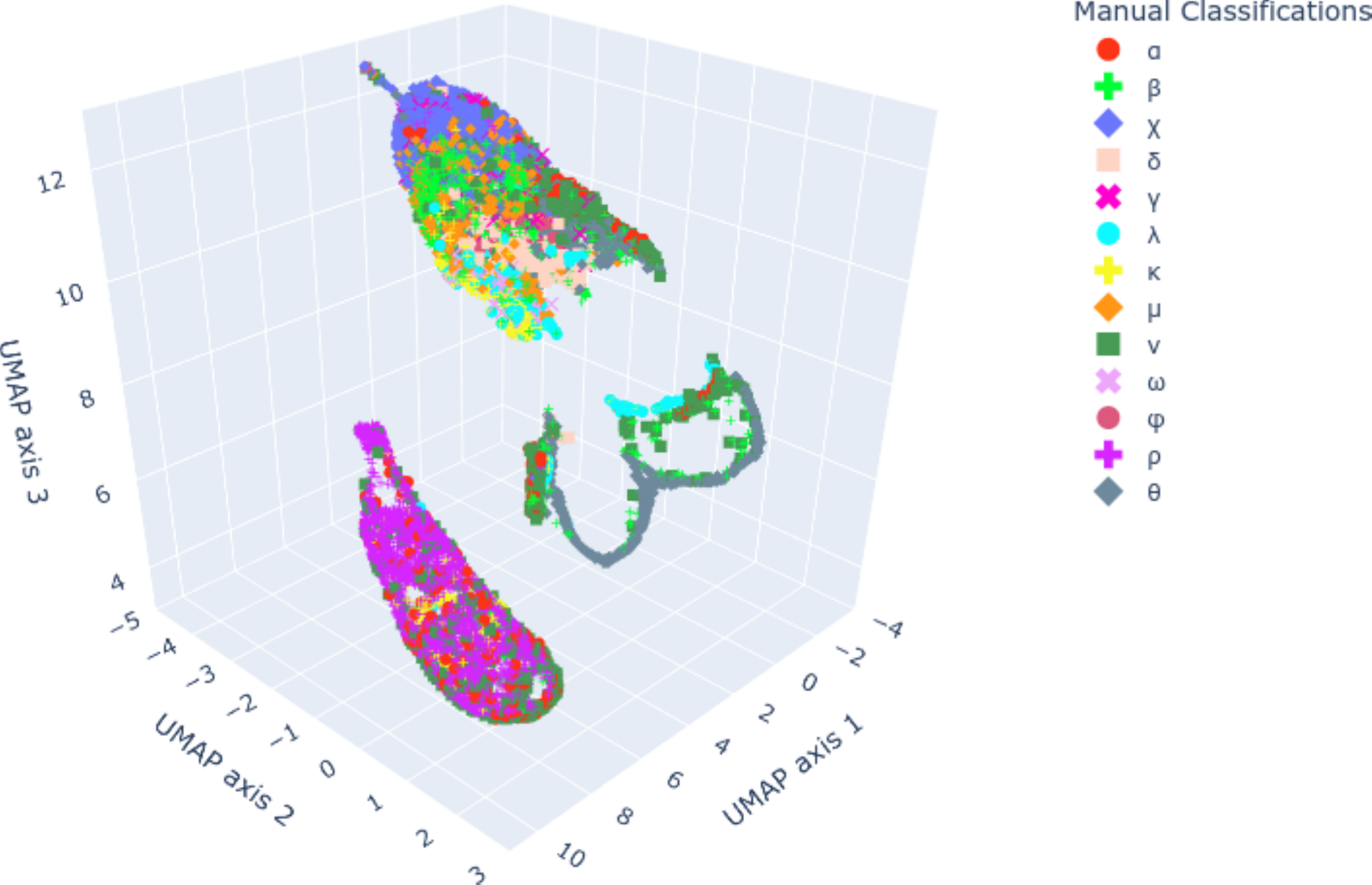}
    \caption{A UMAP projection of the latent variables for 256 second long segments. There are 34327 classified segments from the data set that are plotted, each of which are represented by a single point in this graph. Each segment is plotted as a colored shape according to its class which were manually classified. We label the three visible clusters as follows: left is the "horn", right is the "flag" and top is the "brain".}
    \label{fig:umap_proj}
\end{figure*}

\begin{figure*}
    \centering
    \includegraphics[width=0.8\textwidth,height=0.4\textheight]{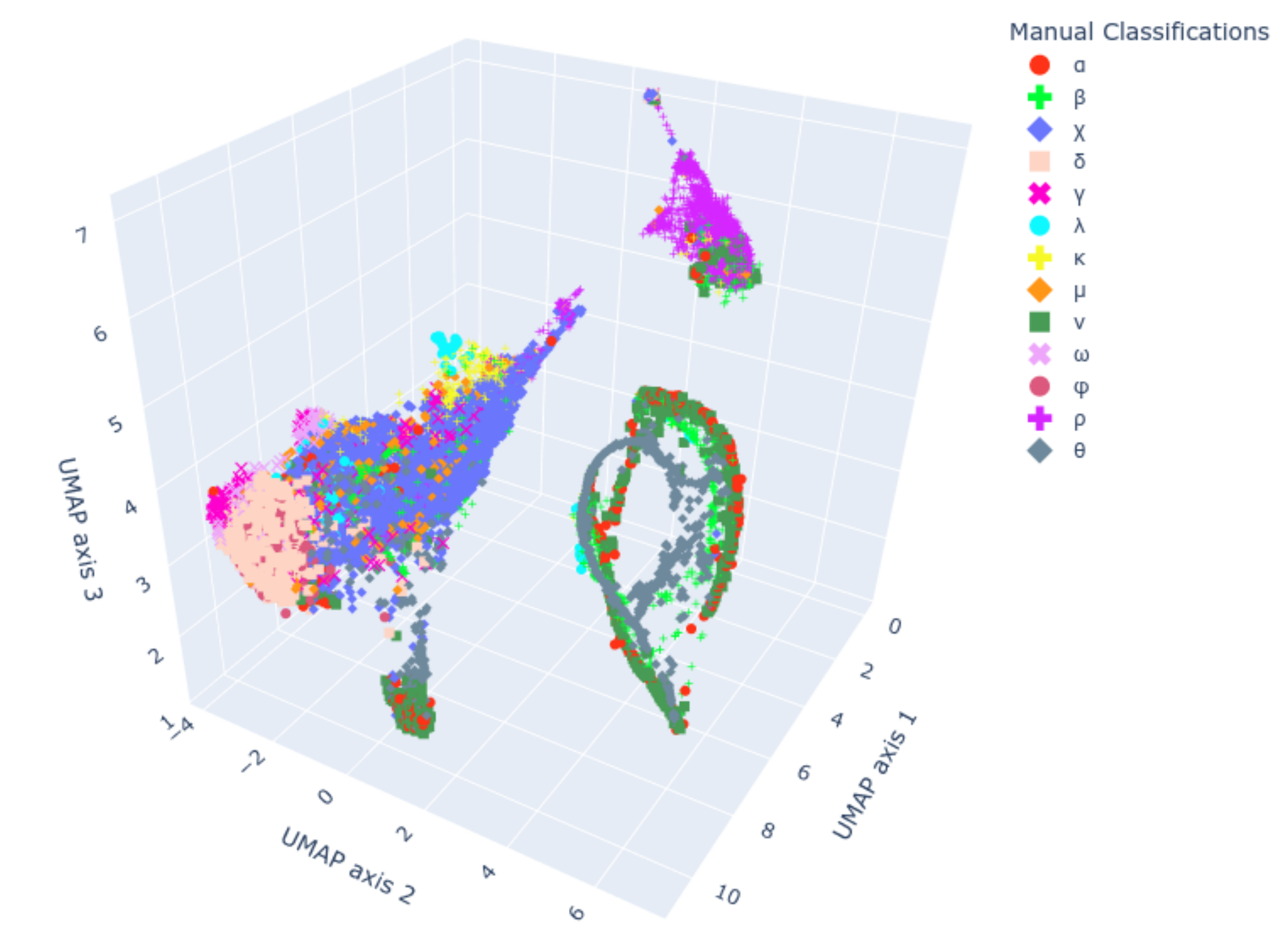}
    \caption{A UMAP projection of the latent variables for 1024 second long segments. Each segment is plotted as a colored shape according to its class which were manually classified. Notably, structures formed in Fig. \ref{fig:umap_proj} are still present in this projection, albeit rotated, flipped or distorted.}
    \label{fig:umap_1024}
\end{figure*}
We utilised the AdamW optimiser \citep{loshchilov2017} from Pytorch for training of the network. We used a learning rate of $10^{-4}$ and betas: $\beta_1 = 0.95$ and $\beta_2 = 0.999$. We used the default epsilon rate of $10^{-8}$ (a term added to improve numerical stability) but used a weight decay rate of $0$. These parameters were chosen by human testing, informed by observed performance. We did not perform a hyper-parameter search as we could not afford the computational cost required. We investigated but ultimately did not adopt the amsgrad variant of the algorithm \citep{reddi2019} as we found there to be no meaningful gains to the network's learning. We continually trained the network until training had been performed over the whole data set 50 times consecutively without improvement. We performed minor adjustments to these parameters to attempt further reduction in loss but only negligible improvements were achieved.

We also attempted training with the stochastic gradient descent with momentum (SGD + momentum) optimiser \citep{sutskever2013} from Pytorch but found minimal to no benefits over use of AdamW while SGD+momentum required more oversight in choice of parameters. This is unsurprising considering AdamW also does stochastic descent and utilises momentum as well.

The final mean reconstruction loss (as specified in Eq.~\ref{eq:wmse}) across the whole data set was $34.98$ for intensity light curves and $3.49$ for both of the color ratio curves (a breakdown of reconstruction loss by class and data channel can be found in Table \ref{tab:reconstruction_errs}). Thus, when taking into account that reconstruction loss is calculated with respect to the error on each of these values (see eq. \ref{eq:wmse}), the color ratio curves would appear to be fitted more accurately, but the disparity in loss between the different components is due to the large disparity in relative error between the components.

A decision to stop training was made when we were able to reliably minimize the loss across the three components and that samples of reconstruction across different behaviours were considered satisfactory. We deemed reconstruction satisfactory when the network was able to grasp prominent behaviour that were particularly characteristic of GRS 1915: for example, the intense peaks of the $\rho$ class. We took the extra case of looking at samples as well as minimizing loss as we found that the smallest values of loss over the whole data set did not necessarily correspond to the most accurate grasps of behaviour across the whole data set. We additionally considered the derivative of loss over each batch of training. If there was no improvement in loss in 50 epochs, we stopped training.

\subsection{Using UMAP for latent variable interpretation} \label{clustering}
\begin{figure}
    \centering
    \includegraphics[width=\linewidth,height = 0.3\paperwidth]{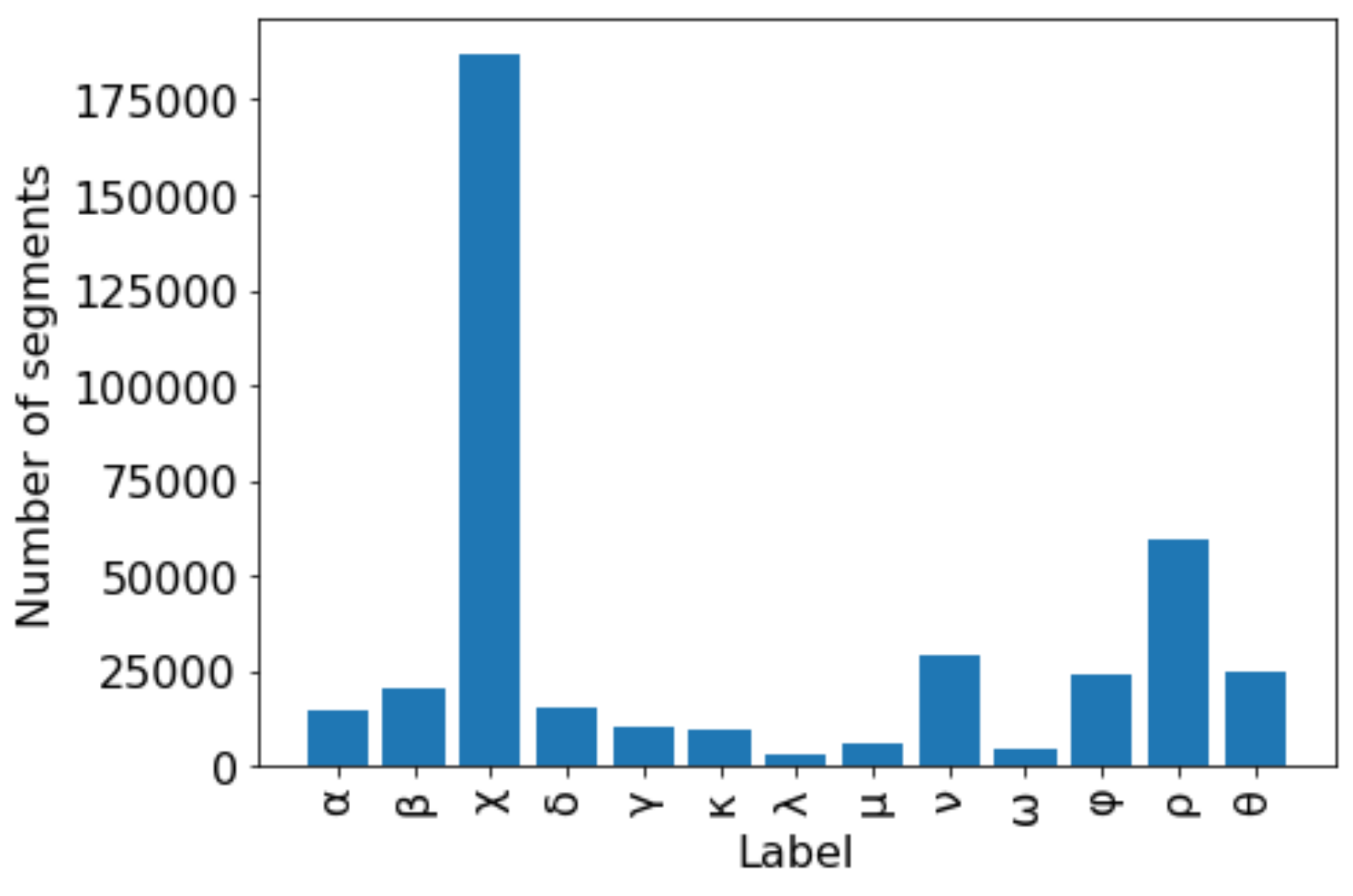}
    \caption{\myRed{A histogram of the number of segments of each class for the manually labeled data. The data set is clearly very imbalanced, with class $\chi$ outnumbering all other classes by a large margin and several classes such as $\lambda$ and $\omega$ represented by only a few examples.}}
    \label{fig:hist_classes}
\end{figure}

We visualised clustering based on the latent variables produced by the auto-encoder. We utilised a technique called Uniform Manifold Approximation Projection (UMAP) \citep{McInnes2018} to create a 3-dimensional projection of the latent variables of each data segment. UMAP aims to preserve the global structure of high dimensional data in lower dimensions. We chose to project into 3 dimensions rather than 2 as we found there to be critical information for interpretation in the 3-dimensions that was lost when plotting only in 2 dimensions. 

While UMAP performs better at preserving global data structures than other dimensional reduction techniques such as Principal Component Analysis (PCA) \citep{Yang2004} and t-distributed stochastic neighbor embedding (t-SNE) \citep{vandermaaten2008}, distances in this projection are still not directly physically interpretable. Distances between points in the projection can be used to understand the structure of the data in the projection, but it cannot be used to directly make judgements of the physical phenomena behind the data.

We show the projection in Fig. \ref{fig:umap_proj}, plotting the 256 segments with increments of 64 seconds. We plot only observations that have been manually labeled and unambiguously classified. These clearly defined classes numbered 1035 out of the overall 2412 observations. We remind that all data sets - not merely those with unambiguous classification -  were used in the construction of the UMAP projection, as the behaviour of intermediate or transitional observations are also critical in mapping GRS 1915's behaviour. Although a minority of the data by observation count, the classified observations represent $79\%$ of the data by time, and a corresponding majority of the 256s segments.

\section{Results} \label{Results}

Fig. \ref{fig:hist_classes} shows the distribution of segments by manually labeled class. Below, we discuss the properties of the various classes in turn, through the lens of our encoded projection maps. 

We created two UMAP projections: one for latent variables from 256 second long segments and another for latent variables from 1024 second long segments (discussed in greater detail in section \ref{1024}). Fig. \ref{fig:umap_proj} shows the UMAP projection of all manually classified 256 second long light curve segments. Fig. \ref{fig:umap_1024} shows the UMAP projection for all manually classified 1024 second long light curve segments. Each colored point on the projection represents a different segment and is colored by its classification. We note to the reader that the structures in both Fig. \ref{fig:umap_proj} and Fig. \ref{fig:umap_1024} are in fact the same structures but orientated differently with slight distortions. In this section, we primarily focus on analysis of the 256 second long segments' UMAP. We find the UMAP (Fig. \ref{fig:umap_proj}) separates the 256 second long segments into 3 distinct clusters. These clusters are made up of 3 different types of events: the ``brain" at the top contains noisy, small relative changes in intensity but otherwise featureless light curves, the ``horn" on the bottom left contains light curves with intense repeating flaring and the ``flag" on the bottom right contains long intensity rises and decays which either precede or follow a large increase or drop in intensity.

Of particular note is the layout of flaring segments within the horn. The horn is hollow such that the points that make up the horn are purely contained within a surface. The horn's length determines the period of flaring - the higher the point on the horn, the shorter period of the flare. The phase of the flaring is encapsulated in the circumference of the horn - the same observation at a later time segment appears at a different part of the circumference, assuming the observation's flares do not vary in period.

\subsection{Light curve classes} \label{UMAPres}
We outline our findings on each of the 13 classes in the 256 second timescale below:

\subsubsection{$\alpha$}
\begin{figure*}
    \centering
    \includegraphics[height=0.25\textheight]{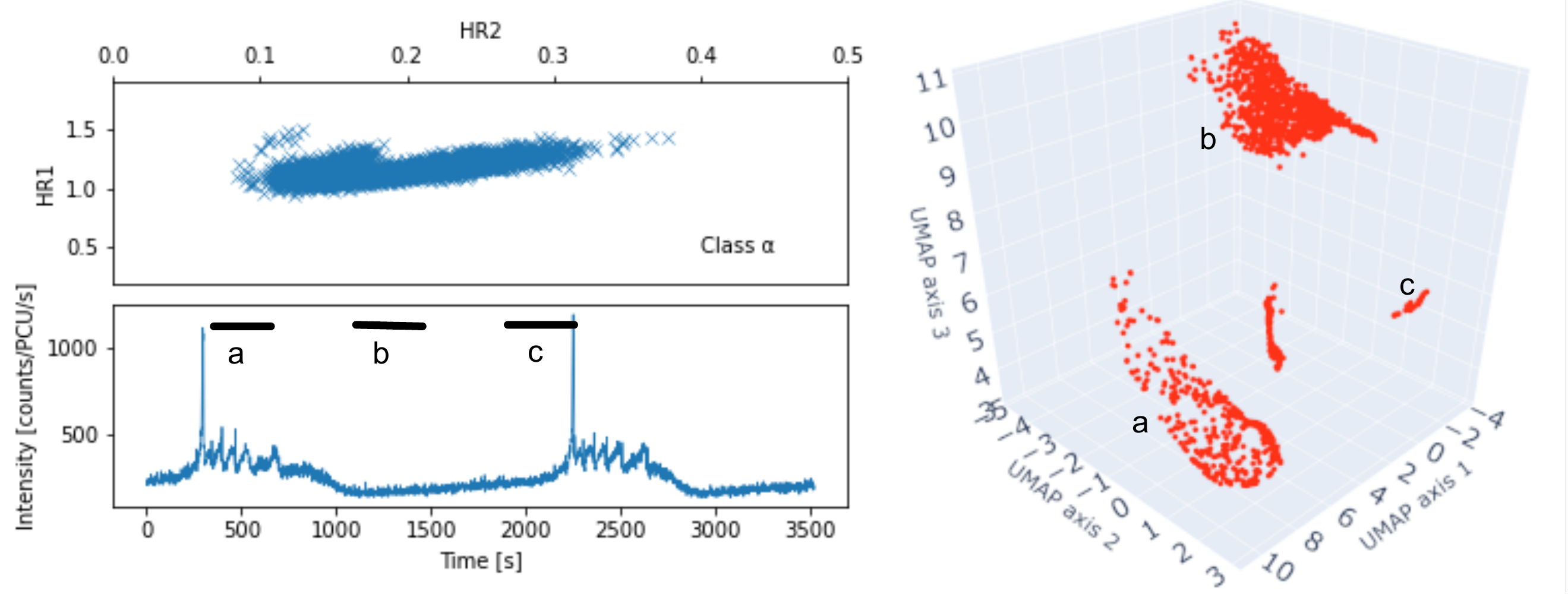}
    \caption{\myRed{Left: An example of the $\alpha$ class. The example light curve (lower left panel) shows a small increase from low activity behaviour prior to a very high amplitude flare. This flare is followed by approximately 300 seconds of lower intensity flaring. After this lower intensity flaring ends, we see a gradual decline in intensity over 200-300 seconds. This pattern then repeats. The colour-colour diagram (upper left panel) covers the middle of the diagram. $\alpha$ is generally harder than other classes, with activity in the class proportionally increasing its higher energy emission leading to a diagonal increase in HR1 and HR2 in the colour-colour diagram. We have added three black bars over the sections of the time series, indicating their approximate length. The letter attached to each segment is used to highlight membership of a particular island in the UMAP projection on the right. As can be seen, the rapid flaring of segment “a” corresponds to the “horn” island, also associated with rapid flaring in other classes, the rather quiet segment “b” corresponds to the “brain” region, where other less variable classes like $\chi$ also live, and finally segment “c” corresponds to a small section of the “flag” island associated with rising trends. Right: $\alpha$'s distribution in the UMAP.}}
    \label{fig:alpha}
\end{figure*}

Fig. \ref{fig:alpha}, right, depicts the distribution of $\alpha$. This class is distributed across all three clusters. $\alpha$ (as defined by B00) features long ($~1000$s) quiet periods, followed by a strong flare and a few 100s of oscillations. These oscillations have a time scale of a few dozen seconds and become progressively longer. These patterns repeat in a very regular manner. The network splits segments into all 3 clusters. 

The horn contains a variety of different shaped flaring segments, going from as low as 2 flares to as many as 5 flares. The color distribution shows a constrained diagonal where each of the colors increases in magnitude at a similar rate.
    
The flag cluster contains 2 clear split mini-clusters of $\alpha$: the left mini-cluster contains light curves with slow decreases in intensity and the right mini-cluster contains light curves with slow rises in intensity, punctuated with a single, extremely high intensity flare. The color distributions of these clusters are an outlier among $\alpha$: we observe a "two-pronged" shape when a flare is observed in the segment. This is associated with an increase in hardness.
    
The brain contains mostly low-intensity, relatively flat change light curves which are dominated by red noise and a small cluster of very small rising light curves which occupy the small thin thread in the distribution contained within the brain. Upon inspection, this small mini-cluster of rising light curves is a result of the network failing to understand and recreate the behaviour of this sample of light curves.

\subsubsection{$\beta$}

\begin{figure*}
    \centering
    \includegraphics[height=0.25\textheight]{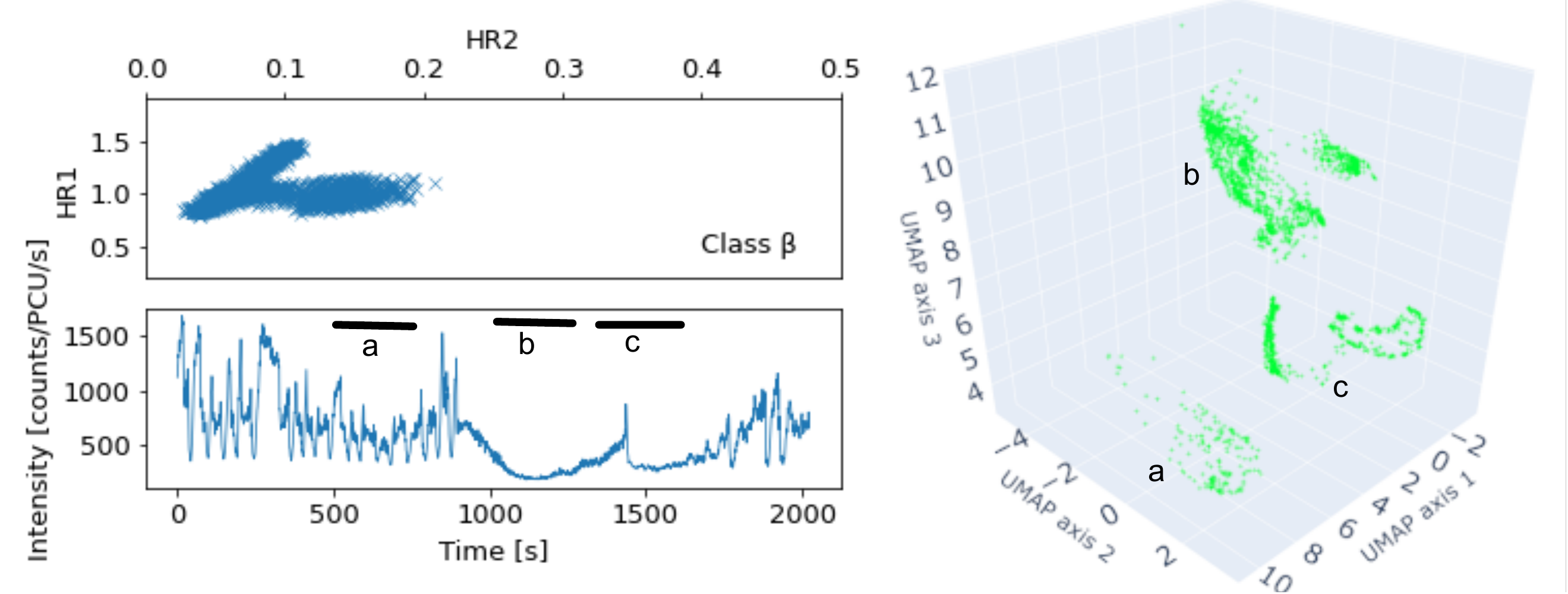}
    \caption{\myRed{Left: An example of the $\beta$ class. The example light curve (lower left panel) shows erratic flaring that lasts for 800 seconds that stops at a moderately high intensity. The light curve then drops in intensity over the course of approximately 200 seconds before slowly increasing again over 300 seconds. This slow increase is punctuated by a short, singular flare which is followed by low intensity activity. This low intensity activity then transitions back to the higher intensity erratic flaring, similar to the start of the light curve. The colour-colour diagram (upper left panel) is constrained to the top left of the colour-colour diagram. It also features a "wish bone" shape, composed of two intersecting arms. In one diagonal arm, we see activity increase proportionally in both 5-13 keV and 13-60 keV. In the second horizontal arm, we see 13-60 keV emission relatively increase while 5-13 keV emission stays relatively constant. We have added three black bars over sections of the light curve indicating their approximate length. The letter attached to each segment is used to highlight membership of a particular island in the UMAP projection on the right. The rapidly flaring segment "a" corresponds to the "horn" island, the quieter segment "b" corresponds to the "brain" region, and segment "c" corresponds to the "flag" island associated with rising and falling trends. Right: $\beta$'s distribution in the UMAP.}}
    \label{fig:beta}
\end{figure*}

Fig. \ref{fig:beta} depicts the $\beta$ class. $\beta$ occupies what appears at first glance, a similar UMAP space to that of $\alpha$. Its primary similarity is its distribution across all three clusters. A notable difference however is the distribution of $\beta$ vs. $\alpha$ across these clusters.

The horn and the brain both contain the characteristic flaring portion of the segments. The natural assumption would be that these flaring features would be mapped exclusively to the ``horn" but the network splits the segments into two groups depending on the extremity of change in the color ratios. Another explanation may be that the network is unable to recreate particularly short time period features leading the network to be unable to differentiate a large number of the flaring segments from noise. Where the segments differentiate the most is in the color-color distribution. The horn contains more more modest variation in colors and exhibits very little to no structure. The brain contains the "prong" shapes in the color-color distribution (visible in Fig. \ref{fig:beta}). This is characterised by an area of low HR2 associated with the full extent of HR1 creating one prong and an area of increasing HR2 associated with proportionally increasing HR1.

\begin{figure*}
    \centering
    \includegraphics[height=0.25\textheight]{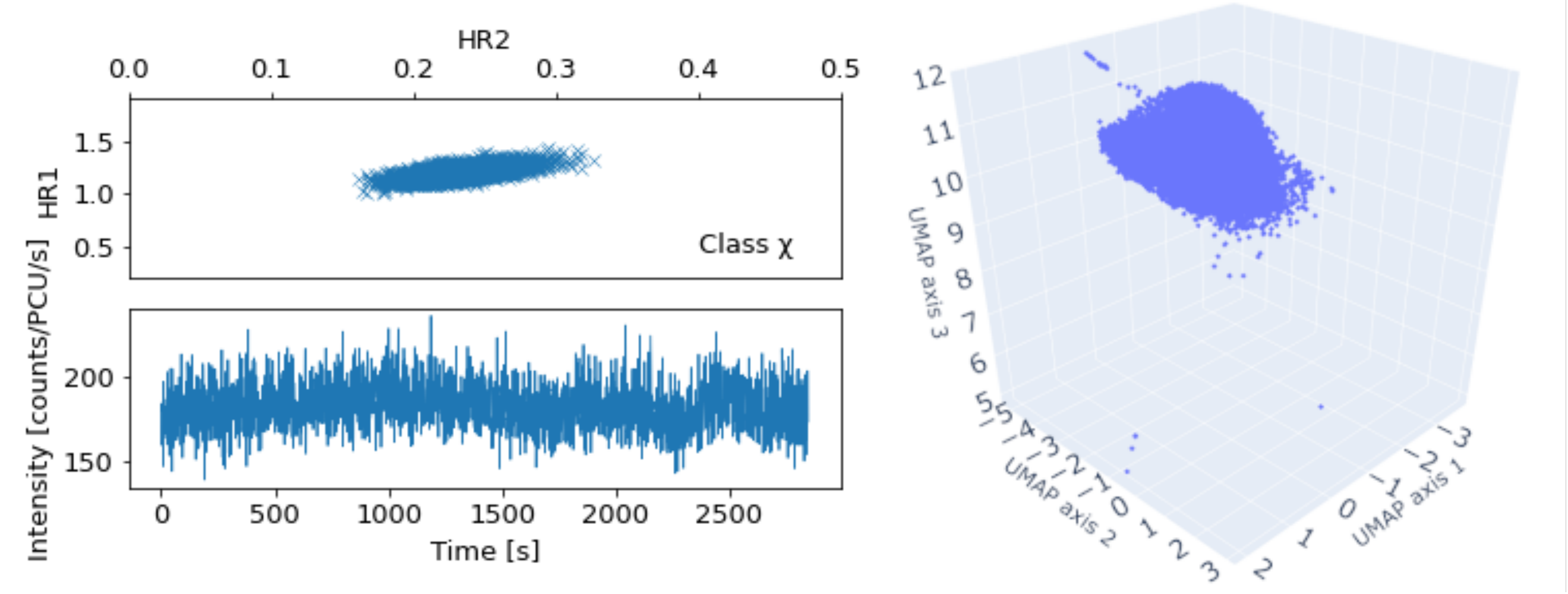}
    \caption{\myRed{Left: An example of the $\chi$ class. The example light curve shows rapid variability without clear patterns (lower left panel) and the colour-colour diagram is loosely clustered in the centre of the diagram (upper left panel), showing little structure. Right: $\chi$'s distribution in the UMAP. $\chi$ is constrained to the "brain" cluster with other non-structured classes.}}
    \label{fig:chi}
\end{figure*}

\begin{figure*}
    \centering
    \includegraphics[height=0.25\textheight]{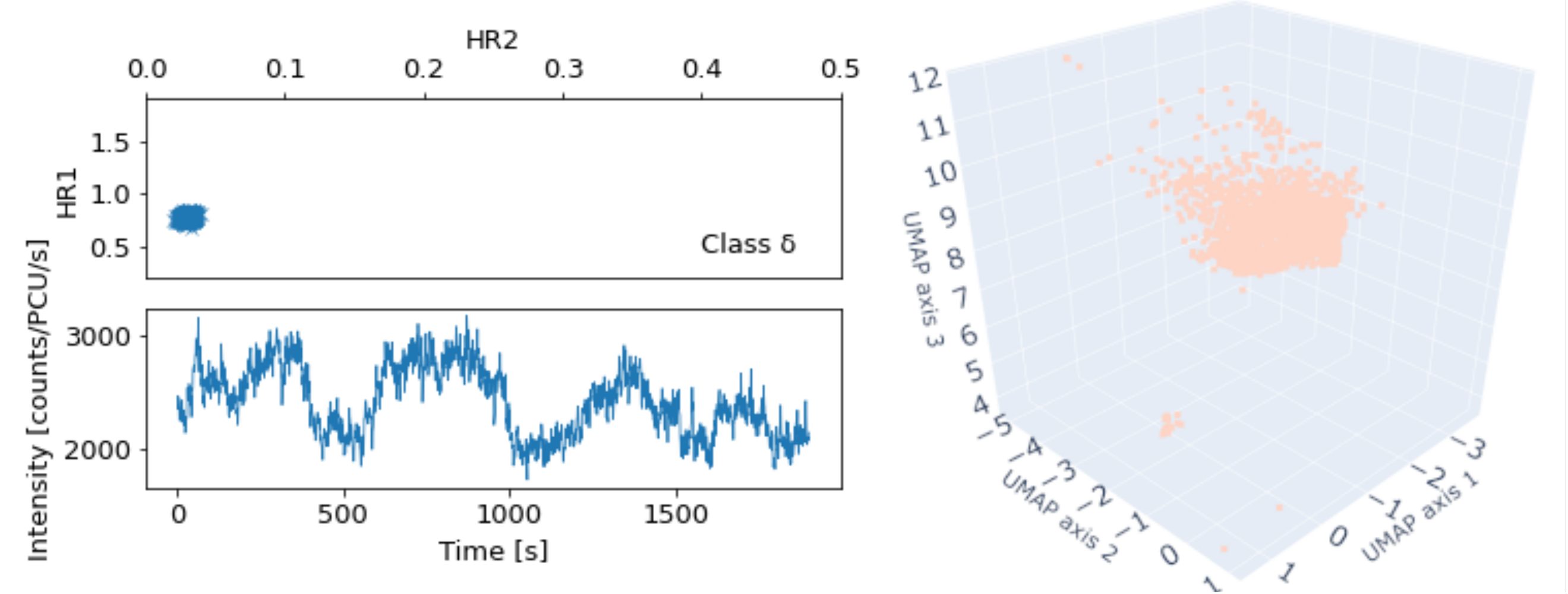}
    \caption{\myRed{Left: An example of the $\delta$ class. The example light curve (lower left panel) shows rapid variability with some structure but no clear trends or periodicity. The colour-colour diagram is tightly constrained to the lower left of the diagram (upper left panel). Right: $\delta$'s distribution in the UMAP. $\delta$ is constrained to the "brain" cluster, with other examples of classes with rapid variability but little clear structure.}}
    \label{fig:delta}
\end{figure*}

The flag contains the long stretches of intensity rise and decay as well as the single large flare characteristic to the $\beta$ class. Where flaring is prominent, we observe the same "pronged" effect in the color-color distribution seen in the brain. Segments of intensity rise and decay have very little structure in the color-color distribution. These segments generally can be described as having a Gaussian distribution in color.

\subsubsection{$\chi$}

Fig. \ref{fig:chi} depicts the $\chi$ class. $\chi$ is by far the most frequently observed class. Due to its very consistent structure, it is confined to one cluster. Color variation is constrained in $\chi$ and shows little structure. Both colors vary linearly with each other.

\subsubsection{$\delta$}

Fig. \ref{fig:delta} depicts the $\delta$ class. It is far more confined than the classes of $\theta$ or $\lambda$. Its color-color distribution is particularly constrained, showing almost no structure at all with the exception of occasional points of high hardness associated with a drop in intensity. These can be explained by a drop in lower energy emission with no associated drop at higher energy levels.

As such, it is not particularly surprising to see it primarily occupy the brain cluster. $\delta$ primarily overlaps in the UMAP with $\phi$, $\mu$, $\chi$, and $\gamma$.

\subsubsection{$\gamma$}
\begin{figure*}
    \centering
    \includegraphics[height=0.25\textheight]{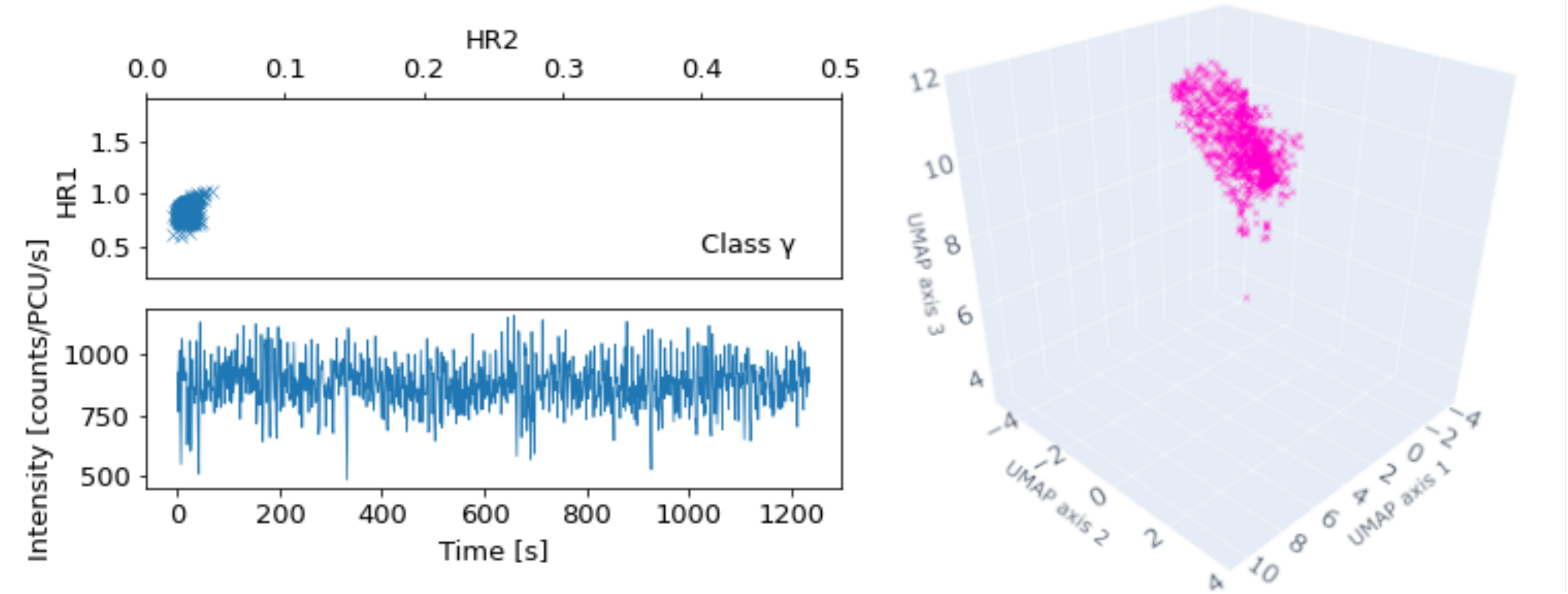}
    \caption{\myRed{Left: An example of the $\gamma$ class. The example light curve (lower left panel) shows very rapid variability, with little structure. The variability does include sharp drops in intensity, distinct from other classes. The colour-colour diagram (upper left panel) is tightly constrained to the lower left of the diagram, not dissimilar to that of $\delta$. Right: $\gamma$'s distribution in the UMAP. $\gamma$ is constrained to the "brain" cluster, similar to classes such as $\chi$ and $\delta$.}}
    \label{fig:gamma}
\end{figure*}
Fig. \ref{fig:gamma} depicts the $\gamma$ class. $\gamma$ is almost indistinguishable from $\chi$ to the network. The network's insensitivity to absolute intensity values is the main cause of this as well as its inability to capture the short, sharp drops in intensity characteristic to this class.

It also overlaps heavily in UMAP space with $\delta$. $\gamma$ is considerably more variable than $\delta$ in its color-color distribution but lacks structure as it generally shows a linear trend between colors.

\subsubsection{$\lambda$}
\begin{figure*}
    \centering
    \includegraphics[height=0.25\textheight]{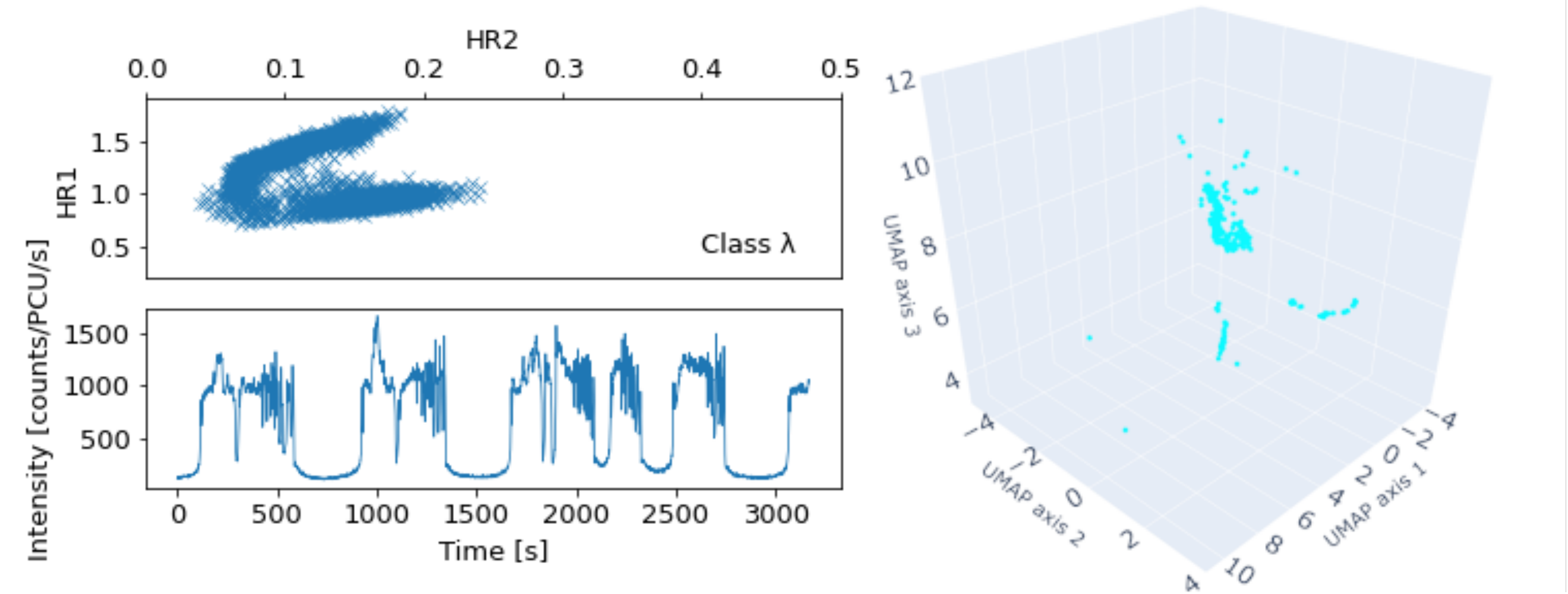}
    \caption{\myRed{Left: An example of the $\lambda$ class. The example light curve (lower left panel) shows drastic transitions from low variability, low intensity behaviour to high variability, high intensity behaviour. In particular, the high intensity behaviour varies in the length of activity, lasting between 100 to 500 seconds. The colour-colour diagram (upper left panel) spans almost the upper left side of the diagram, shaped in a curve not dissimilar to a wishbone. As the light curve varies in intensity, it travels along the curve in the colour-colour diagram to be dominated by soft and hard emissions. Right: $\lambda$'s distribution in the UMAP. $\lambda$ is infrequently observed so represents only a small portion of the data set but we see it constrained both to the brain when in periods of low activity and constrained to the flag in periods of high activity.}}
    \label{fig:lambda}
\end{figure*}
Fig. \ref{fig:lambda} depicts the $\lambda$ class. $\lambda$ makes up the smallest amount of segments over the whole data set, making up only $\sim 0.6\%$ of the data. It occupies two clusters: the brain and the flag. 

Similar to $\alpha$, the segments in the flag are split into two mini clusters. These clusters principally are dependent on the slope of the light curve: the left mini-cluster contains light curves that are decaying in intensity over time while the right contains those growing in intensity. This separation of mini-clusters is amplified by the color-color distributions. Both clusters contain segments with prong color-color distributions. The left mini-cluster contains a more equal distribution of points over both prongs. The right mini-cluster contains more data in the softer prong.

The brain contains light curves with sustained high intensity with semi-frequent intensity drops. The network likely differentiated these from the flaring segments due to how long the light curve remains at high intensity. Particularly of note is the color-color distribution shape. While primarily forming the more common prong shape, some segments have a distribution similar to a full color loop.

\subsubsection{$\kappa$}
\begin{figure*}
    \centering
    \includegraphics[height=0.25\textheight]{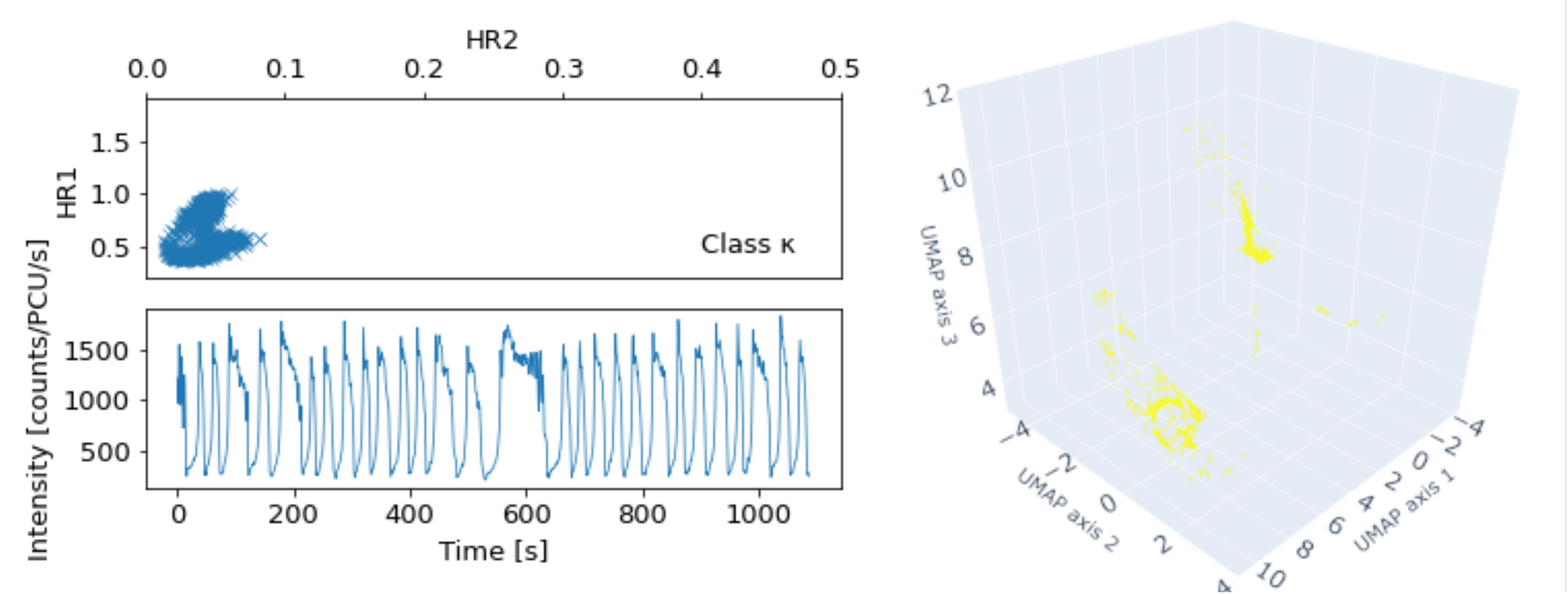}
    \caption{\myRed{Left: An example of the $\kappa$ class. The example light curve (lower left panel) shows extremely rapid flaring that often features a short period of decreasing but still high intensity flaring before dropping to low levels of activity. The colour-colour diagram (upper left panel) is constrained to the lower left corner of the diagram but exhibits a similar "wishbone" shape to that of $\lambda$, albeit with reduced extreme variance of hard emission. Right: $\kappa$'s distribution in the UMAP. $\kappa$ is constrained to both the "horn" island and the "brain" island. Segments constrained to the "brain" island feature particularly rapid flaring on the order of the timing resolution of the data which the network struggles to differentiate from rapid variability with no structure which exists in the $\chi$ class for example.}}
    \label{fig:kappa}
\end{figure*}

\begin{figure*}
    \centering
    \includegraphics[height=0.25\textheight]{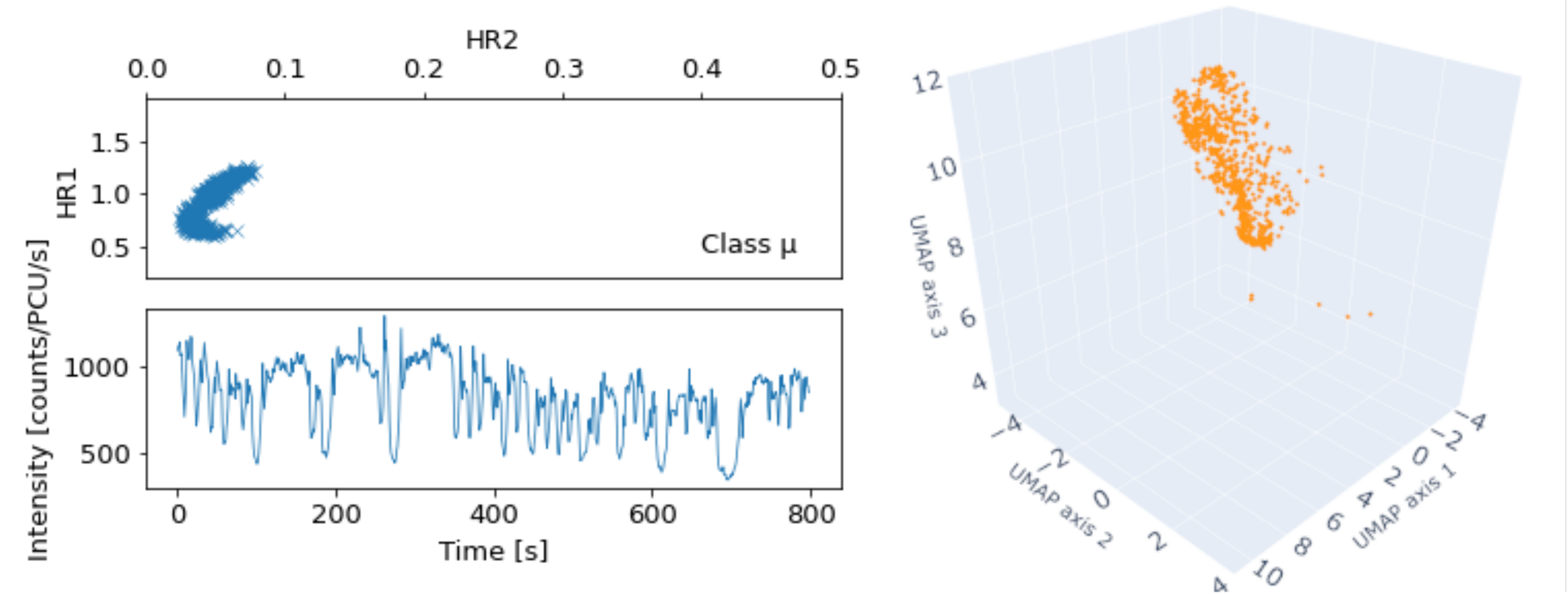}
    \caption{\myRed{Left: An example of the $\mu$ class. The example light curve (lower left panel) shows the light curve rapidly oscillate between a higher intensity and lower intensity where intensity drops for a short period of time before returning for a longer period to higher intensity activity. These oscillations vary in length and depth and show little clear structure on short and long time scales. The colour-colour diagram (upper left panel) is constrained to the left side of the diagram. It also also features a "wishbone" shape but the lower arm is dramatically shorter, indicating that both the harder and softer colour generally increase simultaneously during activity.  Right: $\mu$'s distribution in the UMAP. $\mu$ is constrained to the "brain" island due to its rapid but unstructured variability.}}
    \label{fig:mu}
\end{figure*}

Fig. \ref{fig:kappa} depicts the $\kappa$ class. While $\kappa$ is distributed across all 3 clusters, its inclusion in the brain cluster is plausibly a result of poor reconstructions by the auto-encoder rather than actual meaningful similarities. The potential cause of such consistently inaccurate reconstructions is the short but frequent flaring that occurs in short bursts that is characteristic to $\kappa$. This is borne out by the much higher average reconstruction error that can be seen in table \ref{tab:reconstruction_errs}. Segments in the brain cluster show color behaviour more similar to that of $\chi$, albeit with a more prong-like structure.

A considerable portion of $\kappa$ is confined to the horn cluster. Their location within the horn is greatly dependent on the frequency of flaring in the given segment. High frequency flaring is situated on the tip of the horn. Low frequency flaring (as little as two flares in the 256 second segment) is situated at the base of the horn. Color-color distributions in the horn predominantly appear as full color loops or an intermediary between prongs and loops.

Finally, $\kappa$ segments featuring long periods of reduced intensity, either preceding or following a short, sustained increase in intensity, are situated in the flag cluster. These segments also are differentiated by their color-color distributions. Two areas of color distribution can be seen in these segments: high HR2 with low HR1 values or high HR1 with low HR2 values.

\subsubsection{$\mu$}

Fig. \ref{fig:mu} depicts the $\mu$ class. It is well constrained to the brain cluster. $\mu$ suffers the same poor reconstructions as $\kappa$ with the same likely reasons.

\subsubsection{$\nu$}
\begin{figure*}
    \centering
    \includegraphics[height=0.25\textheight]{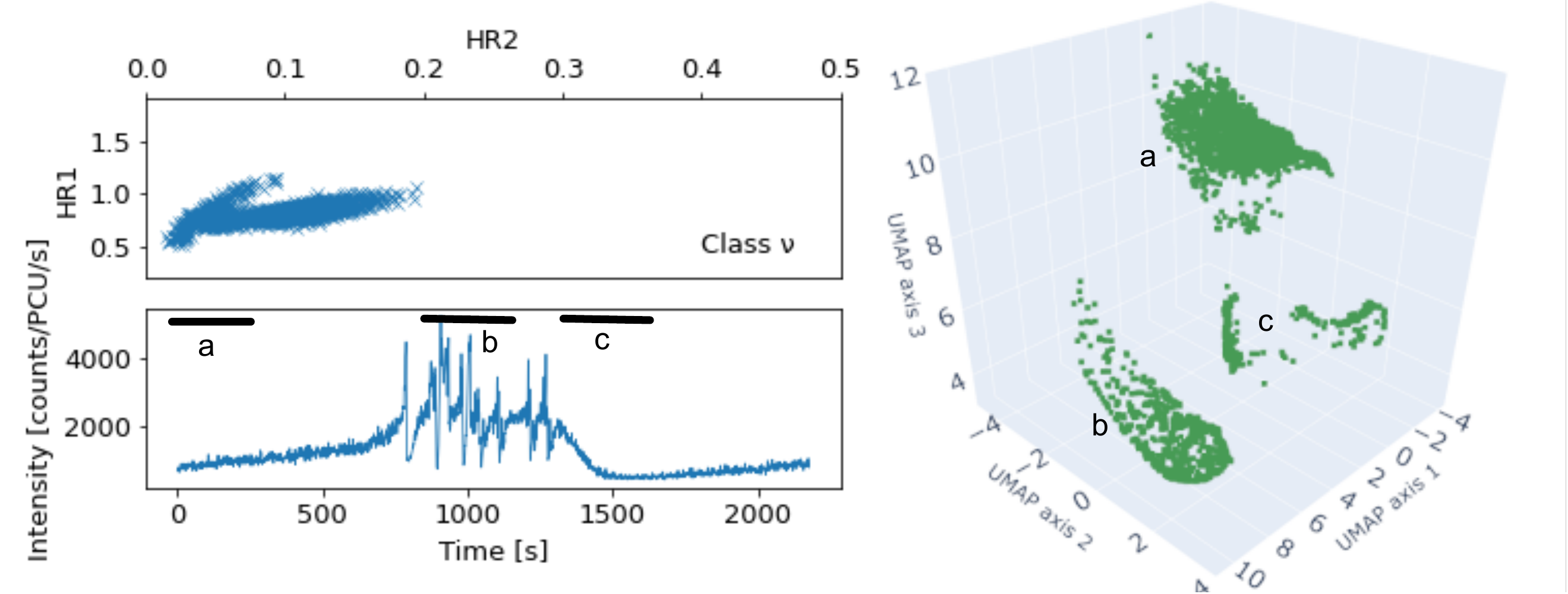}
    \caption{\myRed{Left: An example of the $\nu$ class. The example light curve (lower left panel) shows an initial slow increase in intensity over 750 seconds, leading up to erratic flaring that lasts approximately 500 seconds. Following this erratic flaring, the intensity decreases over 200 seconds before once again slowly increasing as in the beginning of the light curve. The colour-colour diagram extends from the left of the diagram to the center. It features the "wishbone" shape with a longer arm encompassing a large increase in relative 13-60 keV emission while 5-13 keV emission remains proportionally constant to 2-5 keV emission. \myRed{We have added three black bars over sections of the light curve to highlight membership of a particular island in the UMAP on the right. The letter attached to each segment indicates its respective island in the UMAP. Segment "a" is quiet, non-variable activity associated with the "brain" island, sharing it with other less active classes such as $\chi$. Segment "b" exhibits rapid flaring and is associated with the "horn" island. Segment "c" shows a significant decline in activity and is associated with the "flag" island.} Right: $\nu$'s distribution in the UMAP.}}
    \label{fig:nu}
\end{figure*}
Fig. \ref{fig:nu} depicts the $\nu$ class. $\nu$ overlaps with $\alpha$ across all 3 clusters. This strong association reveals structural similarities. The likely similarities include a long decay and then rise of intensity followed by a high intensity flare with a series of following lower intensity flares. The differing length of time that the limit cycles of each of the classes last would imply that it is likely that there are different physical phenomena governing these structural similarities. The confusion between classes is likely due to the relatively short segments that the network is observing, compared to the length of the longer limit cycles for these classes.

\subsubsection{$\omega$}
\begin{figure*}
    \centering
    \includegraphics[height=0.25\textheight]{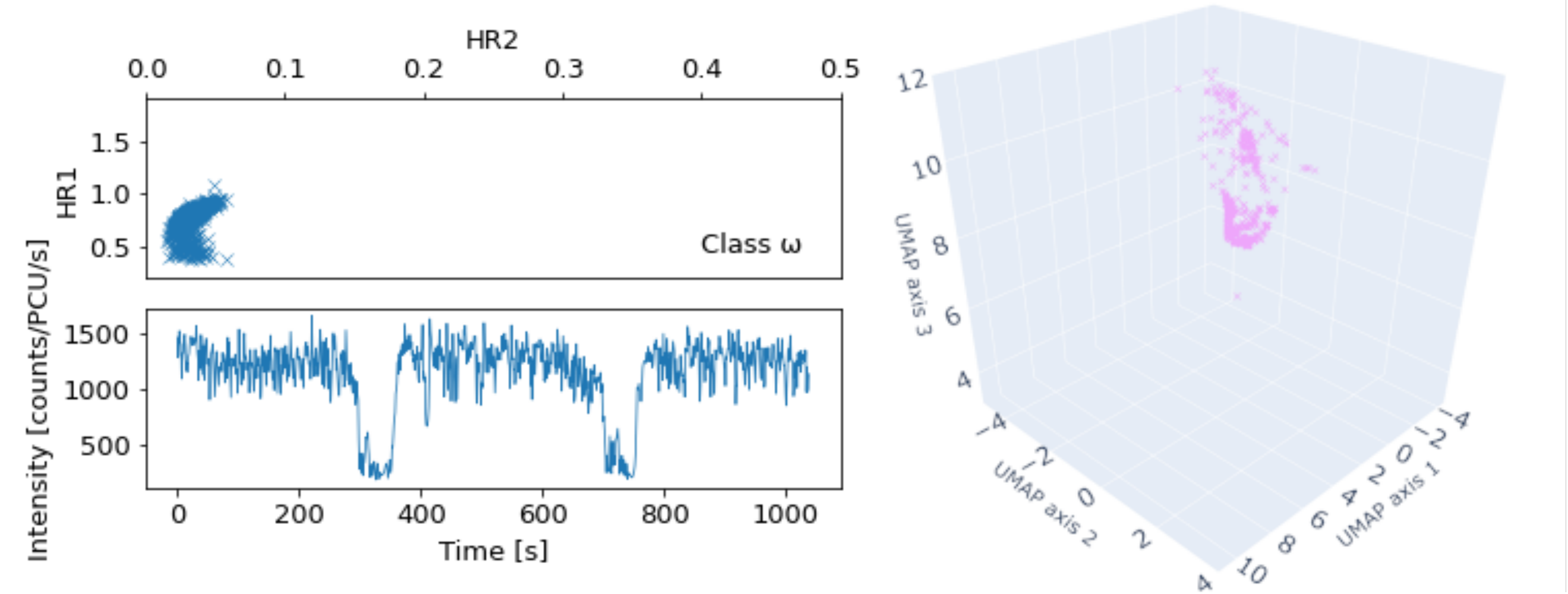}
    \caption{\myRed{Left: An example of the $\omega$ class. The example light curve (lower left panel) shows rapid variability at a higher intensity that is interrupted by severe dips in intensity that last approximately 50 seconds. The colour-colour diagram (upper left panel) is constrained to the lower left of the diagram but forms a clear crescent shape from low HR1 and HR2 values to higher HR1 and marginally higher HR2 values. Right: $\omega$'s distribution in the UMAP.}}
    \label{fig:omega_2}
\end{figure*}
Fig. \ref{fig:omega_2} depicts the $\omega$ class. $\omega$ is the second smallest class of the data set with only 271 segments. $\omega$ is entirely within the brain cluster. Much of the segments heavily overlap with $\chi$, due to the domination of noise over much of the labeled segments.  The rest of $\omega$ overlaps with $\mu$ due to the similar shapes of rapid small dips in intensity within low intensity noise.

There is a small split in the cluster of UMAP values - this is caused by segments of $\omega$ which feature no drops in intensity making them largely indistinguishable from $\chi$. Segments with the characteristic $\omega$ dips have prong color-color distributions while the $\chi$-like segments have unstructured color-color distributions.

\subsubsection{$\phi$}
\begin{figure*}
    \centering
    \includegraphics[height=0.25\textheight]{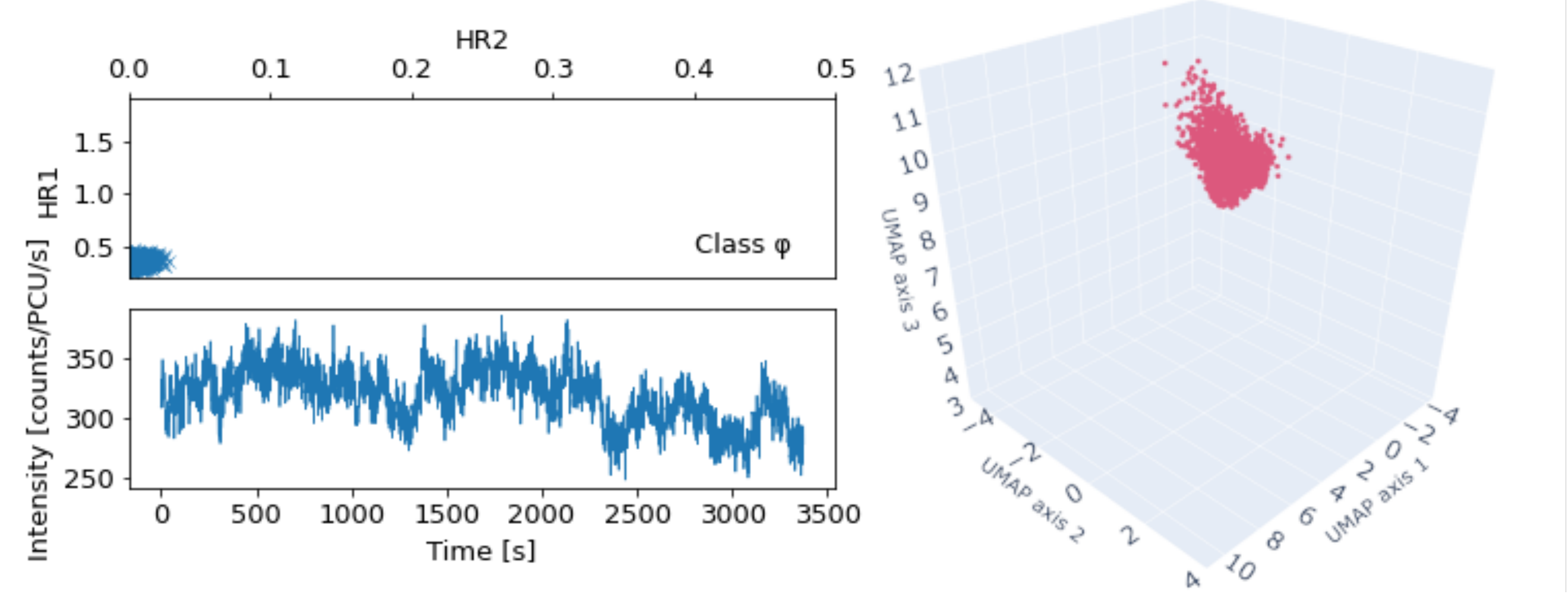}
    \caption{\myRed{Left: An example of the $\phi$ class. The example light curve (lower left panel) shows largely unstructured variability at slightly higher count rates than other unstructured classes such as $\chi$. Most notably, the colour-colour diagram (upper left panel) is constrained entirely to the most bottom left corner, showing very little to no emission in the 13-60 keV x-ray band and lower relative emission by the 5-13 keV band. Right: $\phi$'s distribution in the UMAP.}}
    \label{fig:phi}
\end{figure*}
Fig. \ref{fig:phi} depicts the $\phi$ class. $\phi$ is located in the brain cluster. It heavily overlaps $\chi$, $\delta$, $\gamma$, and portions of $\theta$'s segments. All of these classes feature substantial unstructured noise. 

Of the four overlapping classes, $\phi$ is most strongly correlated with $\delta$. The main difference between $\phi$ and $\chi$ is the much harder color of $\chi$ which is accurately distinguished by the network with $\phi$ appearing as an adjacent ``lobe" to $\chi$. This similarity captured by the network is both intuitive and physically meaningful. These two classes are the only classes that contain no significant variability while maintaining a static state within the color-color diagram.

\subsubsection{$\rho$}
\begin{figure*}
    \centering
    \includegraphics[height=0.25\textheight]{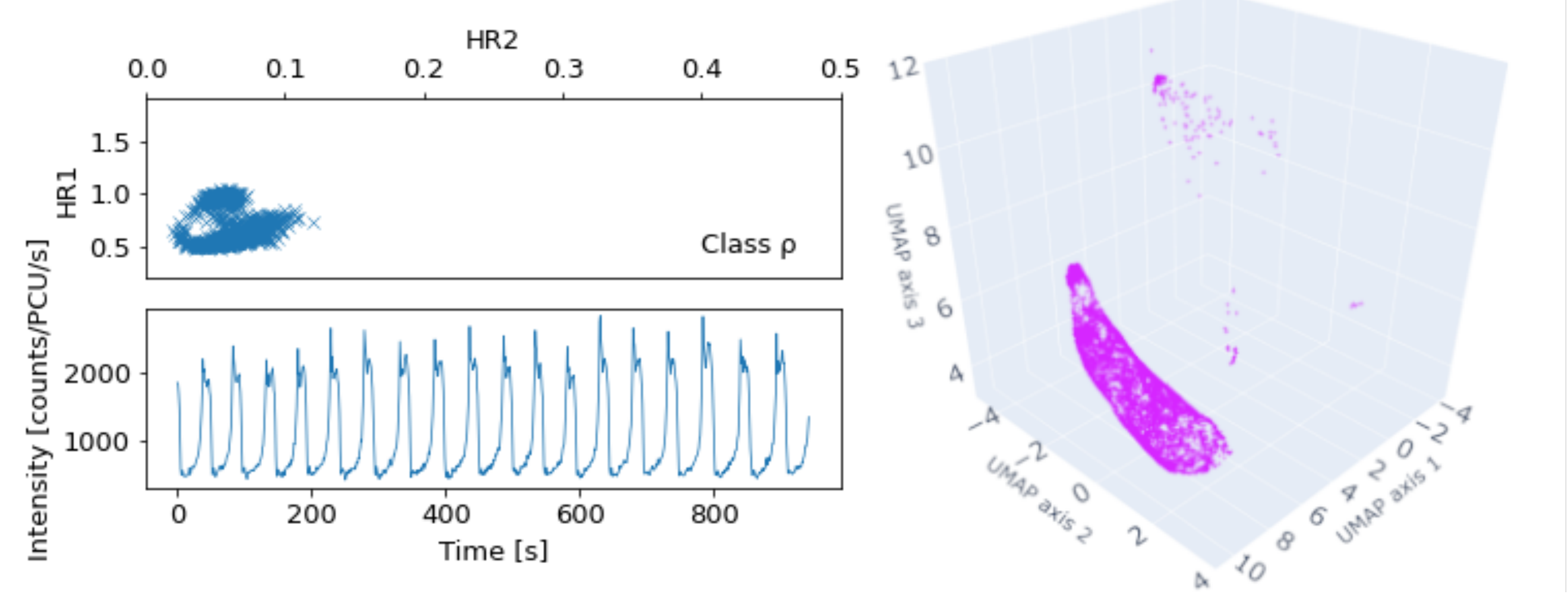}
    \caption{\myRed{Left: An example of the $\rho$ class. The example light curve (lower left panel) shows extremely periodic flaring. Each flare lasts 10-20 seconds and reaches a peak amplitude of approximately 2000-2500 counts/PCU/s. The colour-colour diagram (upper left panel) is constrained to the left of the diagram, showing a "colour loop" where the emission evolves from: being dominated by 2-5 keV emission to mostly dominated by the 13-60 keV emission and then to dominated by the 5-13 keV emission and so forth. Right: $\rho$'s distribution in the UMAP.}}
    \label{fig:rho}
\end{figure*}
\begin{figure*}
    \centering
    \includegraphics[height=0.25\textheight]{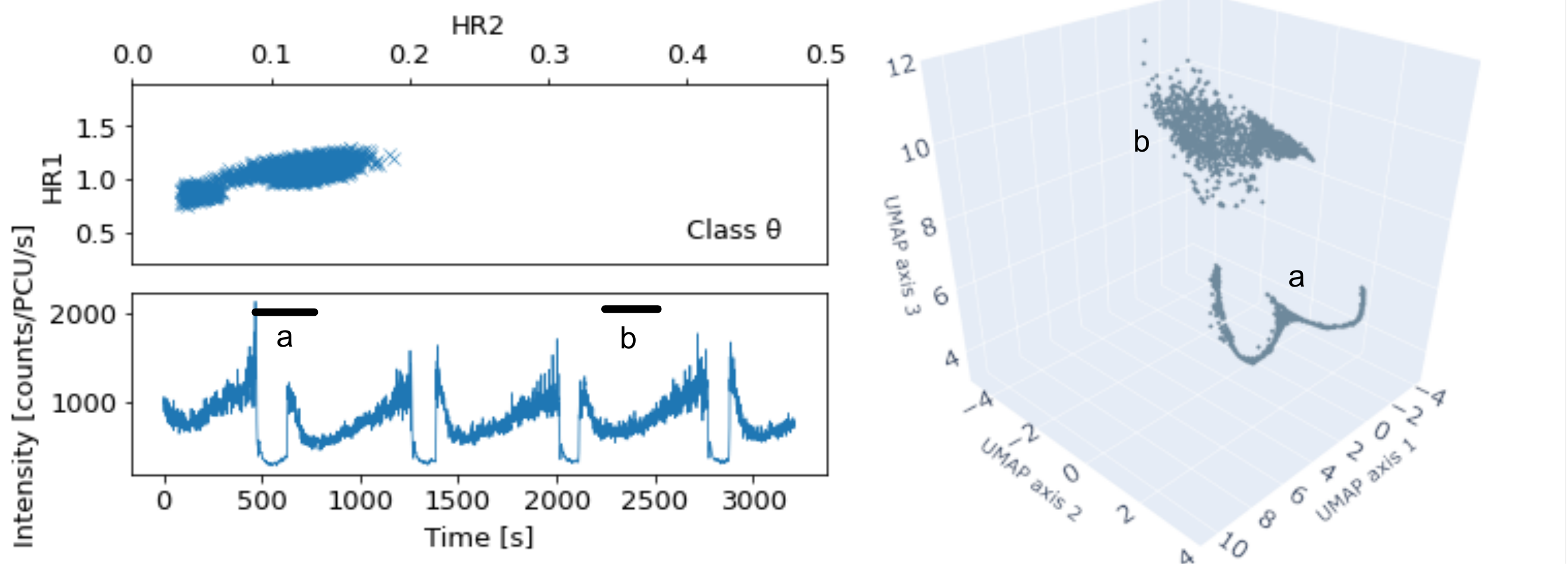}
    \caption{\myRed{Left: An example of the $\theta$ class. The light curve (lower left panel) shows periodic behaviour that can be described as: increases in intensity until a sharp flare followed by a drastic drop in intensity. This drop in intensity lasts for approximately 100 seconds before sharply increasing to approximately 1000 counts/PCU/s. The intensity drops again to approximately 500 counts/PCU/s over a time span of around 50 seconds. This then returns to increasing intensity before a sharp flare once again. The colour-colour diagram (upper left panel) extends from the left middle of the diagram to the center. It shows that 5-13 keV emission stays relatively constant in proportion to the 2-5 keV band (HR1 occupies between 0.75 and 1.25 throughout), but 13-60 keV emission relatively increases considerably over the evolution of the light curve. We have added two labelled black bars over sections of the light curve to indicate their approximate position in the UMAP. Segment "a" indicates a rapid dip in intensity before increasing sharply. This segment corresponds to the "flag" island and its position within the "flag" island is greatly correlated with the position of the dipping behaviour within the segment. Segment "b" indicates a period of low variance activity and is associated with the "brain" island. Right: $\theta$'s distribution in the UMAP.}}
    \label{fig:theta}
\end{figure*}
Fig. \ref{fig:rho} depicts the $\rho$ class. $\rho$ is by far the most unique of all the states and has been subject to particular focus in previous studies and has been nicknamed the "heartbeat" state of GRS 1915 due to its distinctive, consistent flaring in intensity \citep{Neilson2011,Yan2016,Zoghbi2016}. Almost all segments have a color-color diagram exhibiting a color loop. It is also the second largest class after $\chi$ with 3287 segments. 

The network quickly distinguishes $\rho$ as a particularly unique class of observation. Almost all of the $\rho$ segments are constrained to the horn cluster with the exception of mislabeled segments and extremely fast flaring segments. These extremely fast flaring segments have been placed into the brain cluster, likely caused by the network being unable to differentiate the fast flaring from noise dominated light curves.

A few classes do overlap with $\rho$: $\alpha$, $\beta$, $\kappa$, and $\nu$. However, none of these other classes are so exclusively constrained to the horn. $\rho$ remains extremely unique in both human analysis and the network's mapping.

\subsubsection{$\theta$}

Fig. \ref{fig:theta} depicts the $\theta$ class. $\theta$ is also a rather unique class. Its distribution is split between the brain and the flag. Of particular note is that $\theta$ produces the thin connection joining together the two mini clusters that make up the flag. Exploration of this connection reveals a phase change along the length of the connection. The feature change is where a sustained dip in intensity is located within the segment.

\subsection{1024 second timescale} \label{1024}
In addition to the 256 time second segments UMAP that we have discussed at length, we produced another variation of the network and the resulting UMAP (Fig. \ref{fig:umap_1024}) which analyzed the 1024 second long timescale. 

This network was fundamentally very similar to the network we used for the 256 timescale. We only changed two of the pooling layers to more aggressively reduce the size of the data to obtain the same size of code at the end. We found that the UMAP projection structure was broadly the same as our benchmark 256s case and that the issues with modelling persisted at this timescale and the model's reproduction of behaviour was worse as determined by the loss function.

Notably, we found that the vast majority of the same relationships we found between classes in the 256 second timescale is preserved at these longer timescales. This was a surprise as we had expected the classes to further segregate and the transitional cluster to vanish for these longer timescales.

\section{Discussion} \label{discussion}

In this section, we discuss different measures of how the network related together observations within its representation in the UMAP projection.

\subsection{Closeness} \label{closeness}
\begin{figure*}
    \centering
    \includegraphics[height = 0.45\paperwidth]{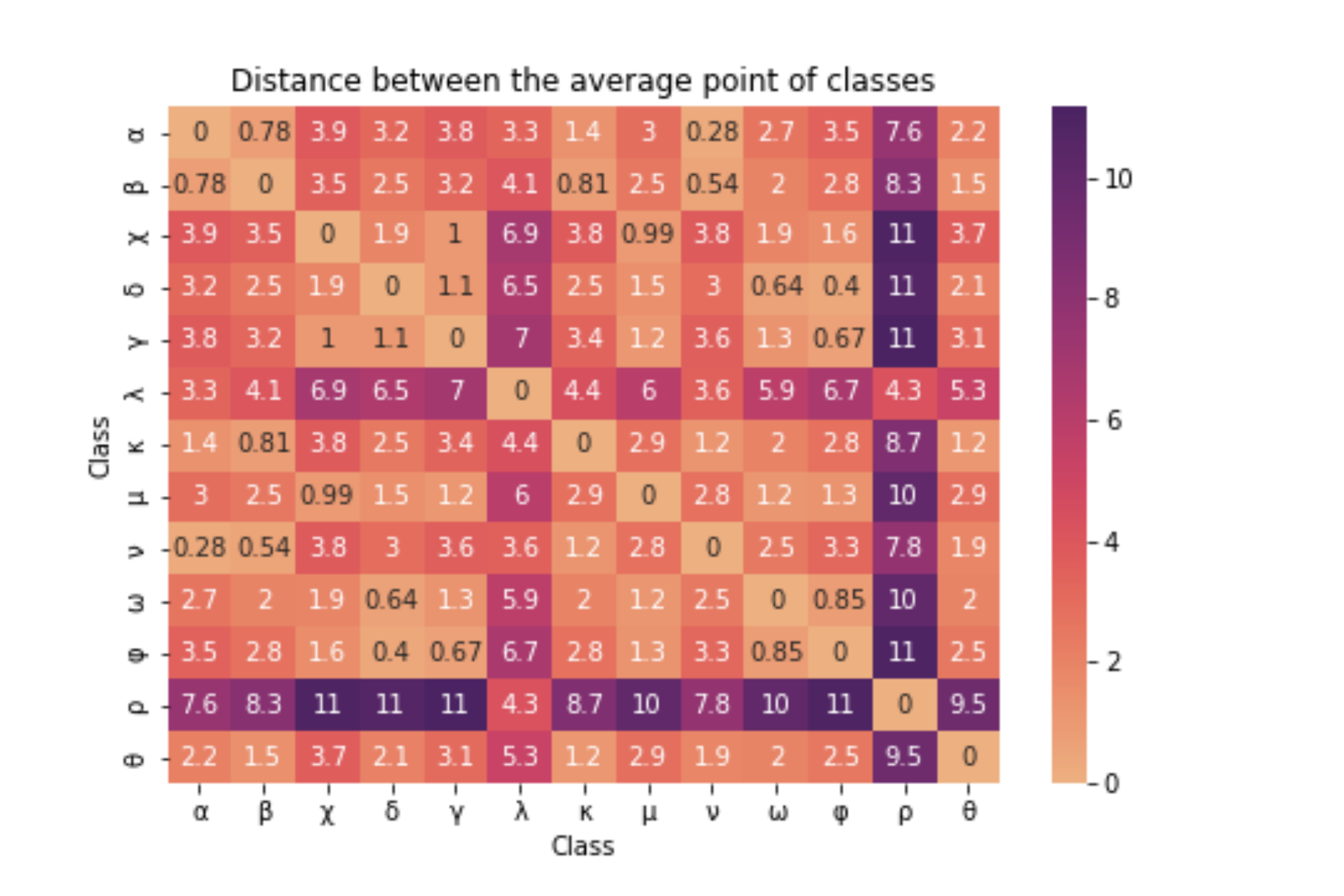}
    \caption{A matrix of the Euclidean distance between the average of the location of points of human labelled classifications in the UMAP projection. One of the representations of how "close" a given class is in distribution to one another. Classes which are a short Euclidean distance from one another have a higher chance of being associated with one another. This method also takes advantage of the clear separation of the UMAP clusters dependent on the type of segment. For classes that are divided between the three clusters, its average location positions itself between the clusters which is then closer to a given cluster based on how many segments appear in a given cluster. This reveals how common the type of behaviour in a given cluster appears within a class.}
    \label{fig:centroid_distances}
\end{figure*}

\begin{figure*}
    \centering
    \includegraphics[height = 0.45\paperwidth]{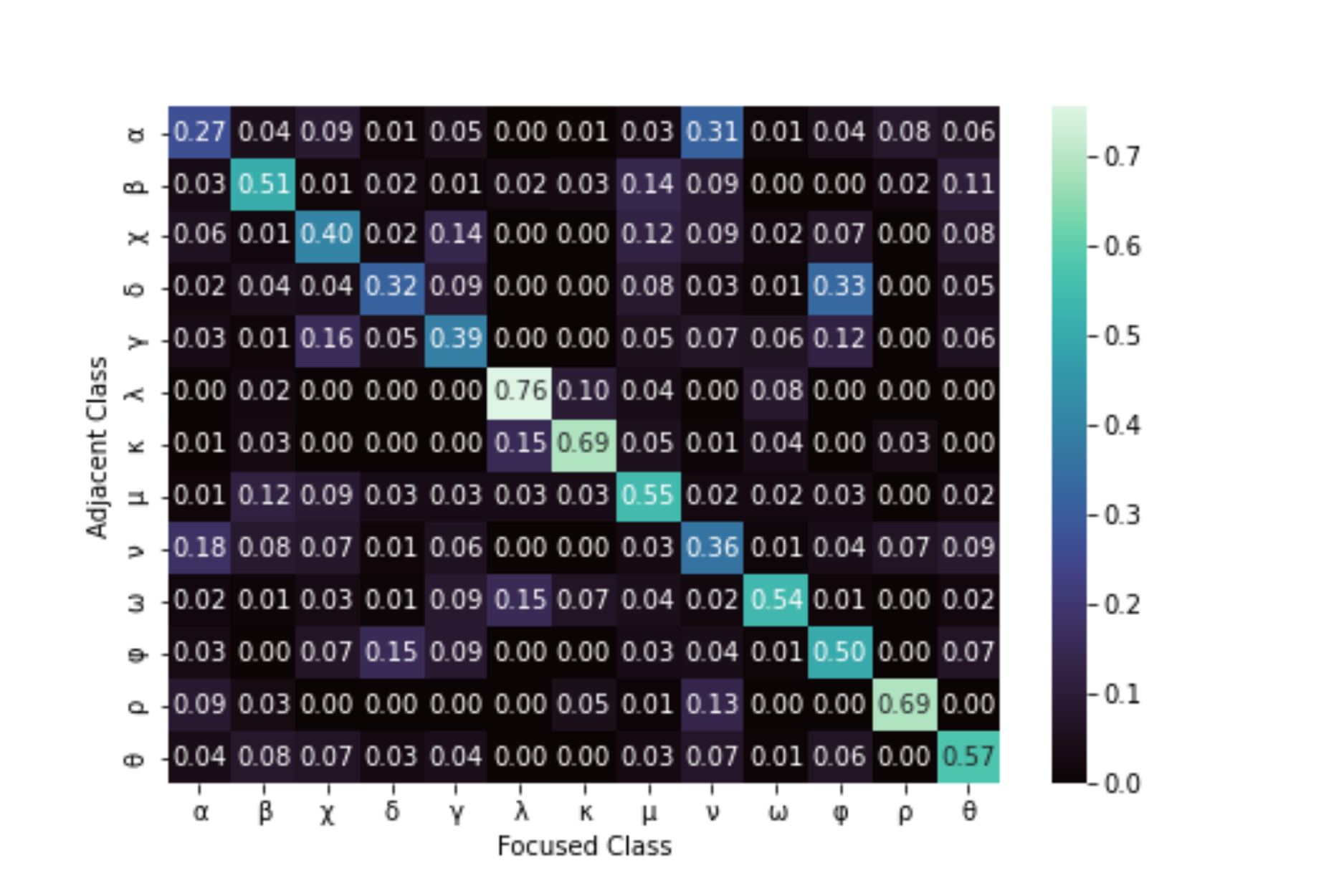}    
    \caption{A matrix showing the nearest neighboring points by class. The matrix is row normalised by the size of sample of the class. The column shows the focused upon class and the row gives the percentage of the given class is the nearest neighbor to the focused upon class. The brighter the square, the more highly associated the two classes are. The data set used to create the UMAP was a balanced form of the original data set specified in section \ref{data_processing}.}
    \label{fig:class_density_neighbors}
\end{figure*}

\begin{figure*}
    \centering
    \includegraphics[width=\linewidth]{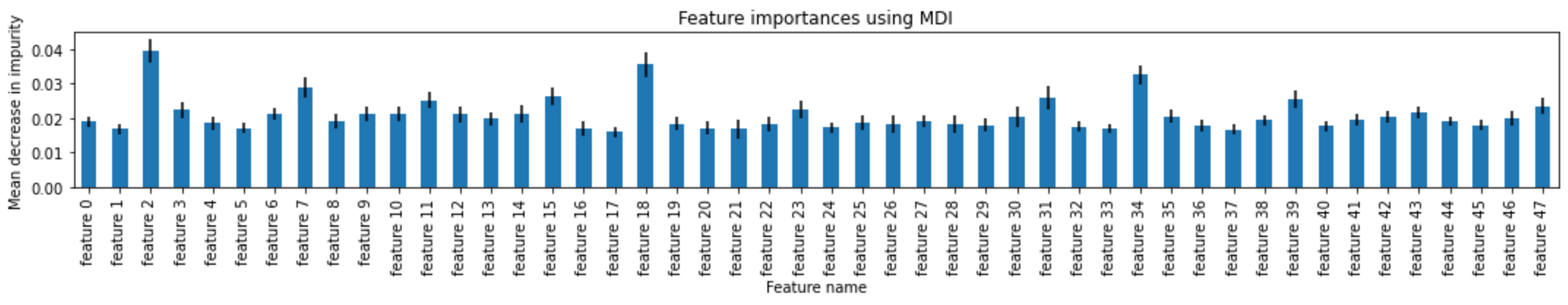}
    \caption{The mean decrease in impurity for each feature within the latent variables for the 256 second segments. Each bar shows the mean decrease in impurity while the standard deviation of the decrease in impurity is shown by the black line at the top of each bar. Note that the 3rd, 15th and 16th features of each information band are most important for defining behaviour.}
    \label{fig:MDI}
\end{figure*}
\begin{figure*}
    \centering
    \includegraphics[width=\linewidth]{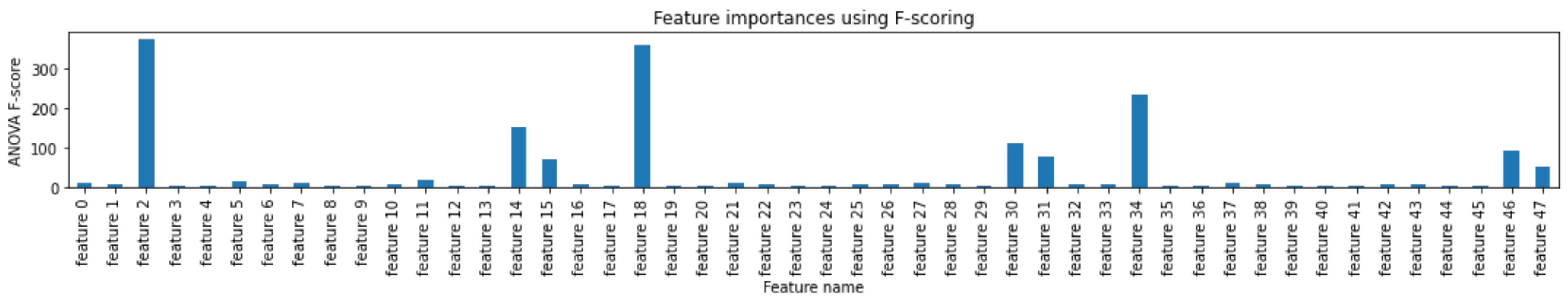}
    \caption{\myRed{The ANOVA F-scores of each feature in a bar chart. A high F-score indicates high degrees of variance between the distribution of the data of different samples. We highlight that there are 9 features that are particularly predictive of a latent vector's class in the B00 system. These 9 features are distributed across the intensity and the two x-ray colours. It should be noted that HR2 is relatively less important than the other two channels of information.}}
    \label{fig:f_scoring}
\end{figure*}

Fig. \ref{fig:centroid_distances} shows a matrix of the Euclidean distances between the average position of each class in the UMAP projection. The purpose of this matrix is to visualise one measure of how ``close" classes of observations are to one another in the view of the network. We noted in section \ref{clustering} that distances in the UMAP are not directly physically interpretable. This is particularly true when considering short distances in the UMAP between individual points. However, when considering aggregate distances as in this case, this is a measure of a general trend in the data and is more justified.

For classes that are divided between the three clusters, its average location positions itself between the clusters which is then closer to a given cluster based on how many segments appear in a given cluster. In our results, we have shown a link between particular types of behaviour and their location to a given cluster. The average point of a class will be closer to a cluster with more segments, and thus the behaviour associated with that cluster is dis-proportionally present.

This line of reasoning can be extended to intra-cluster distributions as well. Even when a class predominately shares and overlaps the same UMAP space as another class, because they spend different proportions of time in the clusters, the distances between their means will generally be substantial. This measure also has the advantage of not being skewed by large sample size classes.

The most visually striking portion of Fig. \ref{fig:centroid_distances} is $\rho$'s distinctly large distance from almost every other class (with the exception of a still moderate distance from $\lambda$). This follows with what we know of $\rho$ already. The proximity of $\alpha$ to $\beta$, $\kappa$, and $\nu$ is also noticeable. This is unsurprising due to all of these classes being distributed across all 3 clusters (owing to the long limit cycles) but this would imply that the relative times spent in the different clusters are fairly similar, even if the behaviours in question are not exactly the same in practice.

Despite $\chi$'s dominant presence in the data set as well as its significant overlap in the UMAP projection with several other classes, its average distance from other classes remains fairly significant in most cases. $\delta$ and $\phi$ also are positioned very close to one another, something noted in earlier discussion.

Fig. \ref{fig:class_density_neighbors} shows an alternative measure of distance between the classes, displaying the proportion of nearest-neighboring data points broken down by class. The matrix has been row normalised. We used a variation of the UMAP plotted in Fig. \ref{fig:umap_proj} which utilized the balanced data set specified in section \ref{data_processing}. This choice was made because otherwise the dominance of $\chi$ over the data set overall will be over-represented in this metric. This measure aims to quantify overlap between classes in UMAP space. This is particularly advantageous in comparing cases where classes may possess very different distributions between clusters such that they are not actually similar in UMAP space but possess a similar average point in the UMAP projection.

The diagonal line of high correlation between a class and itself is reassuring. It shows not only that the B00 system is  robust, but that the network is encoding meaningful features, even in the cases where features vary distinctly over the evolution of each of the classes. 

Combining both Fig. \ref{fig:centroid_distances} and Fig. \ref{fig:class_density_neighbors} allows us to quantify meaningful overlap between classes. We note that $\nu$ and $\alpha$ both have very similar distributions as exhibited by Fig. \ref{fig:centroid_distances} and also a high degree of overlap as exhibited by Fig. \ref{fig:class_density_neighbors}. Futher, we see that the indication of similarity in from the mean-value distance comparison between $\alpha$ and $\beta$ from Fig. 20 is not borne out by Fig. 21. $\phi$ and $\delta$ have considerable overlap in distribution in addition to being closest neighbours, making them candidates for close physical association..

\subsection{Feature importance}
In some machine learning algorithms such as the one used in this paper, the latent variables produced by the network have varying levels of importance in defining and differentiating behaviour. As our data set also contains human classification, a variety of techniques are available to use for defining feature importance within the latent variables produced by the network. We utilized the balanced data set mentioned in section \ref{data_processing} for a portion of the analysis. In the discussion of these latent variables, we refer to the latent variables as a "feature". Features 0-15 refer to portion of the latent variables associated with the intensity model, features 16-31 refer to the portion of the latent variables associated with the HR1 model and features 32-47 refer to portion of the latent variables associated with the HR2 model. We investigated two avenues of assessing feature importance: random forest classifier and ANOVA F-scoring.

\subsubsection{Random forest classifier}
Random forest classifiers \citep{breiman2001} are a technique of classification that utilises random decision trees on different sub-samples of the data set to attempt to classify the sample data based on human defined labels. We utilized scikit-learn's \citep{scikit-learn} RandomForestClassifier module. We used the default settings for the module. This meant that the random forest classifier used the Gini impurity criterion \citep{raileanu2004theoretical} function for measuring the quality of a split as well as having no maximum depth on the random decision trees. Splits in the decision tree only considered a maximum of $\sqrt{n_{\text{features}}}$. Once the random forest classifier is fitted, we can use the mean decrease in impurity (a statistic that defines how often a particular feature was used in creating a split in all of the decision trees in the random forest) to find out which are the most important features for classification.

Fig. \ref{fig:MDI} shows the mean decreases in impurity by feature in a bar chart. Each bar indicates the mean decrease in impurity and the standard deviation is indicated by the thin black line. We see that there are a few features that have a marginally higher mean decrease in impurity - appearing approximately twice as much as the others in the trees. However, there are no features which appear to be completely unimportant as defined by this technique. This would imply that we can't substantially reduce the amount of parameters utilized in the code any further.

\subsubsection{ANOVA F-scoring}
Analysis of variance (ANOVA) F-scoring \citep{st1989} is a statistical technique that utilises sample data mean and distribution of sample data to define the importance of certain features. For the purposes of this paper, it is best understood that a high F-score means that a feature exhibits greater variation between classes. Higher variation between classes is indicative of a feature’s importance.

Fig. \ref{fig:f_scoring} plots the F-score for each feature in the 256 second segments' latent variables. We found every feature to have a p-value of much less than 0.05, determining that every feature had significantly unique distributions for every class that they were important in classification. This is in line with what was determined by the random forest classifier.

There are 9 features that stand out in this technique: 2, 14, 15, 18, 30, 31, 34, 46, and 47. These are essentially the same 3 features that vary the most by class in the code when reconstructing each of: intensity, HR1 and HR2.

Most notably, we find that the importance of the portion of the code that encodes HR2 is less important for determining the class than HR1 or intensity. This is likely because GRS 1915's X-ray signal is strongest in the lower energy bands of RXTE's effective energy range. The behaviour of GRS 1915 in the 13-60keV range evidently does not vary as much between classes as behaviour in the 5-13keV range. This would imply that contemporary instruments such as NICER and Astrosat are very well-suited to obtaining a more precise view of GRS 1915's behaviour. Expanding this technique to the observations taken of GRS 1915 by these instruments would be a good avenue of further research.

\subsubsection{Feature selection}
\begin{figure}
    \centering
    \includegraphics[width=\linewidth]{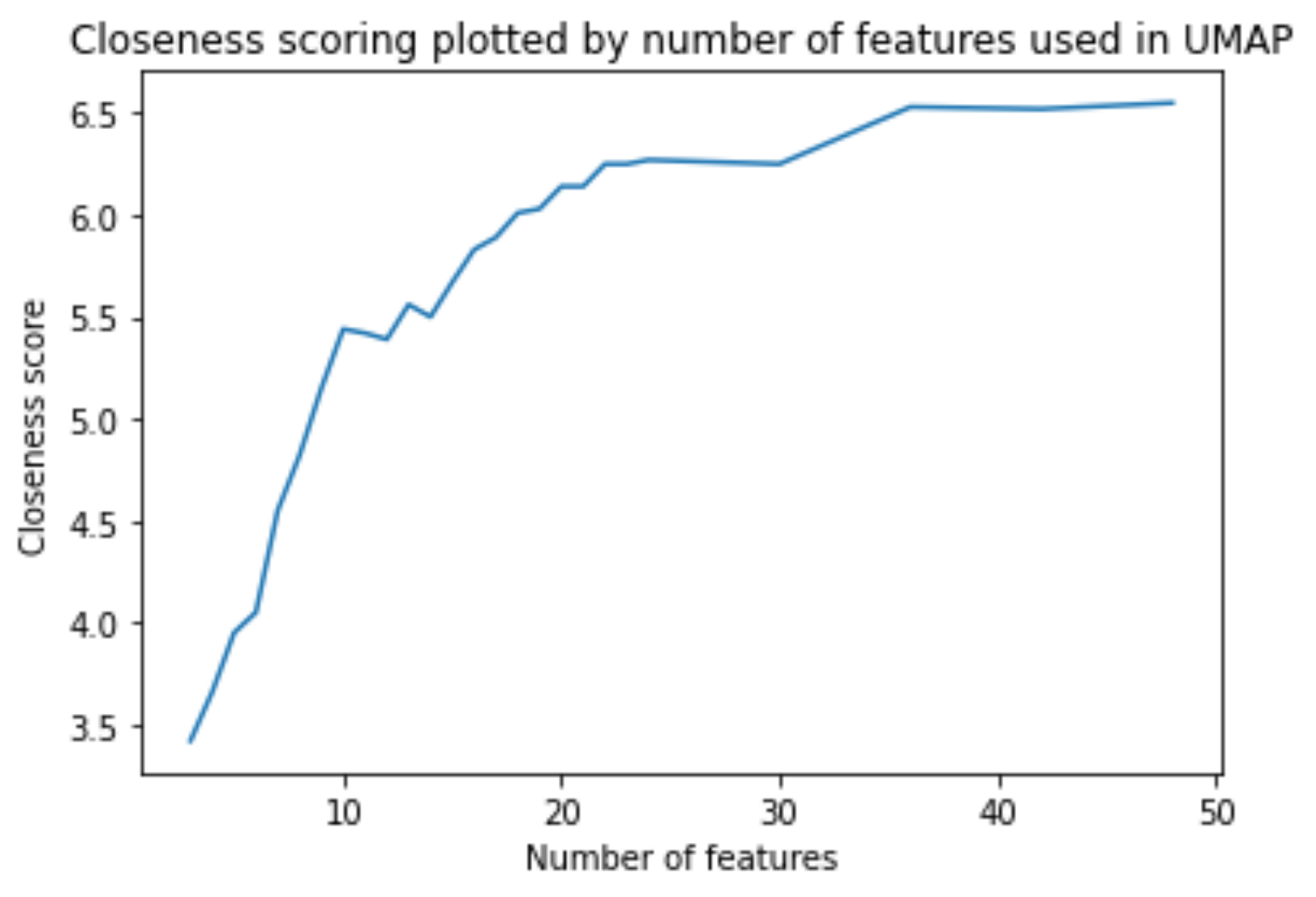}
    \caption{\myRed{Scoring of closeness of all classes. As we add more features, classes in the B00 system become more distinct from one another. We see that the 24 most descriptive features in the latent space describe most of the differences between classes in the B00 system.}}
    \label{fig:closeness_scoring}
\end{figure}

\begin{figure*}
    \centering
    \includegraphics[width=0.49\linewidth]{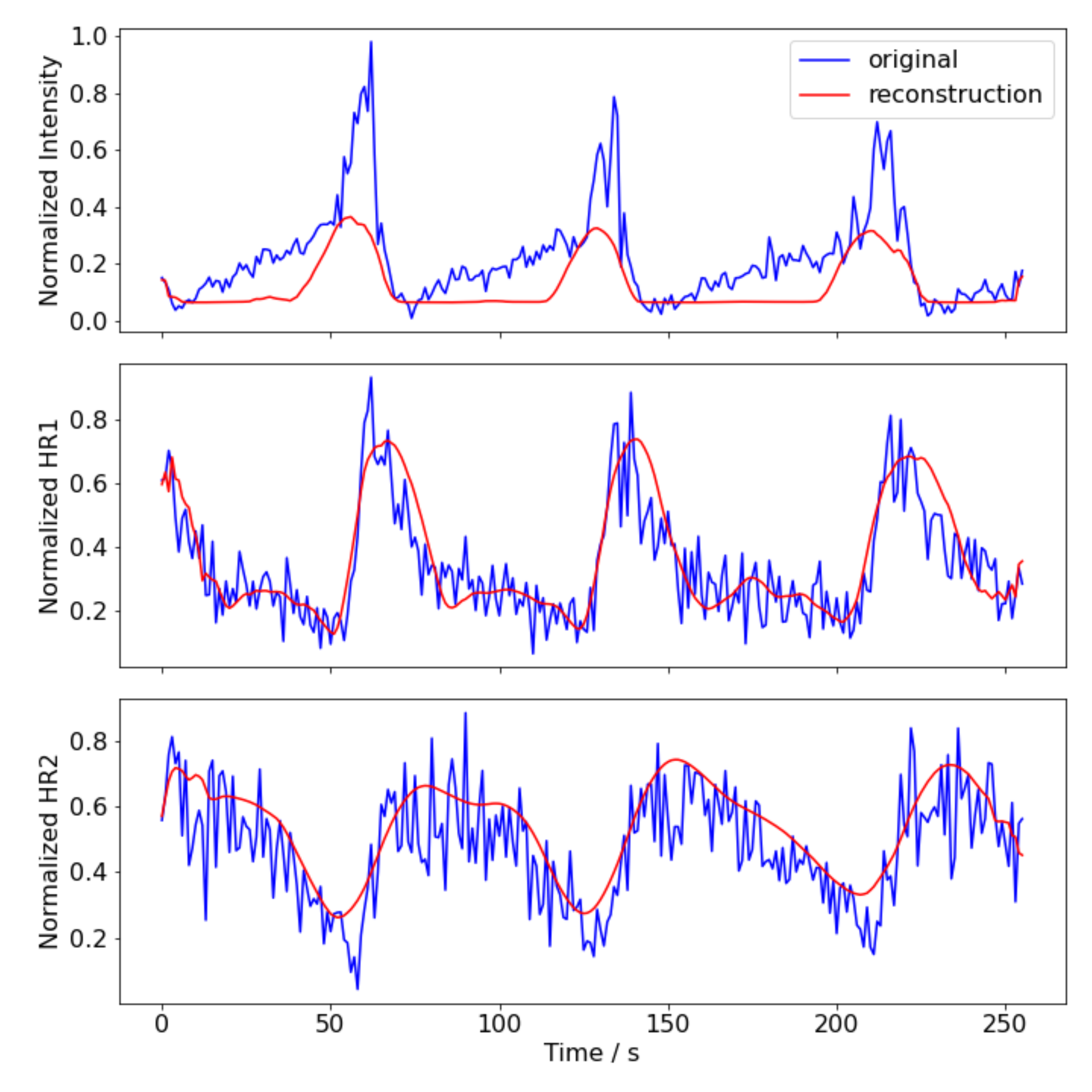}
    \includegraphics[width=0.49\linewidth]{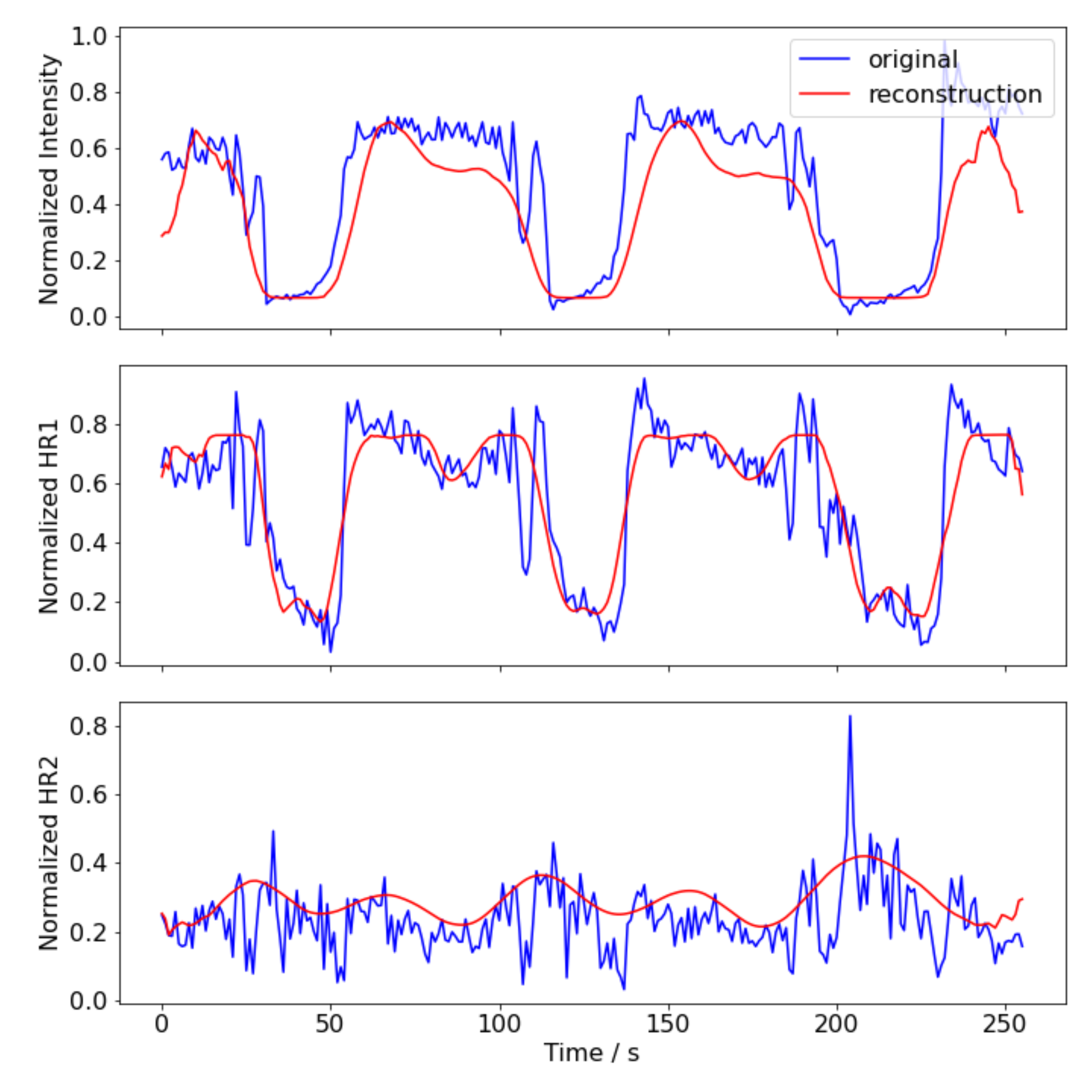}
    \caption{\myRed{Two representative examples of the reconstruction of two segments by the auto-encoder. The top panel shows normalized intensity, the middle panel show normalized HR1 and the bottom panels show normalized HR2. Time is shown on the x axis, shared between the three panels. The left hand observation is an example from the $\rho$ class and the right hand observation is an example from the $\kappa$ class. Each light curve is normalized by mapping the minimum (minus Poisson errors) and maximum value (plus the Poisson errors) of the input to 0 and 1 respectively. While the auto-encoder does not recreate the original observation perfectly, the goal is to capture the general shape of the light curve so these examples have achieved their goal.}}
    \label{fig:reconstruct}
\end{figure*}

By determining feature importance, we can attempt to simplify the model by taking the most important features as determined by ANOVA F-scoring. To quantify the effect of removing features from the latent variables, we can measure a "closeness score". 

We can quantify how accurately the network is placing similar behaviour together by simply adding together each of the class' self-neighboring scores. If each class were to be totally separated, the score would be 13 as each point in the UMAP would have $100\%$ of its neighbors be another of its own class.

Fig. \ref{fig:closeness_scoring} shows how this closeness score varies with the number of features used in the UMAP. To choose which features were used in the UMAP, we chose features in accordance to their F-score. The highest N scoring features were selected to be inputted to the UMAP where N is the number of features. We find that there is a consistent increase in the closeness score with the increase in features until 36 features are included. However, the majority of this increase is contained in the first N=24 features.

\subsection{Comparison to previous work} \label{compare}
Previous studies have focused purely on classification under the B00 system. The work undertaken in this paper extends this to test the assumption that the B00 system is robust and meaningful as well as fully comprehensive of GRS 1915's full extent of behaviour. 

As a comparison in technique, OK22 is much closer to the technique utilised in this paper with its use of an auto-encoder but with key differences in architecture, in particular the use of recurrent neural networks as opposed to convolution neural networks. Recurrent neural networks are more suited to the types of evolution of behaviour seen in GRS1915 as they consider longer term trends and are thus able to achieve better reproduction of behaviour. 

Previous work has also varied in the quantity and variety of information used to inform the machine learning networks. 

H17 utilised considerably more information, opting for manual feature extraction. This included time series features, power spectral features, and color features. A principal component analysis of the power spectrum was built and an auto-regressive model was utilised in the case of the light curve. This allowed for very high accuracy classification for most classes, but required a considerable amount of domain knowledge and large amounts of feature engineering.

OK22 only utilised RXTE's PCA intensity light curves and achieved strong results in classification with fairly minimal information. The auto-encoder in this paper performed autonomous feature extraction similar to that performed in OK22 but also considered the x-ray colors in a time series. This allows for the linking of the color-intensity and color-color diagrams to that of the intensity light curves: key parts of our understanding of black hole accretion states. 

Another considerable difference is that the feature extraction in OK22 is a single step into creating a representation of observations using Gaussian mixture modelling which they called an observation's "fingerprint", while this paper utilizes the latent variables extracted directly and presents them using UMAP for analysis. This direct comparison of latent variables reduces the number of data transformations before analysis. This is particularly important due to how information from the original data is inevitably lost when undergoing each level of compression. This difference in visualisation is suited to different situations. \myRed{The goal of classification in OK22 is to create the fingerprint to allow for comparison of variable length observations and a new observation so as to easily classify new data. This emphasised comparison of full observations, rather than interest in individual segments of the observations themselves. In comparison, this paper aims to compare observations which may not share a class in the B00 system and observe similarities.}

Both OK22 and H17 as well as this paper found levels of confusion between certain classes - OK22 found a high level of confusion between $\omega$ and $\kappa$ while H17 found high levels of confusion of $\gamma$ and $\lambda$ with $\mu$, $\alpha$ with $\xi$, $\chi$ with $\theta$, $\theta$ with $\beta$ and $\rho$ with $\beta$. This paper found the network to confuse together $\alpha$ with $\nu$, as well as $\delta$ with $\phi$. The inconsistencies in confusion of classes between different techniques would seem to imply that the classes may not actually be broadly similar, but only appear so when taking into consideration limited information. Confusion between classes could also be a product of each execution's quirks: the difference in behaviour between classes may be better encapsulated by certain data. Another explanation could be that there is some intrinsic uncertainty for some light curves in terms of their class assignment.

Another notable paper for comparison is \citet{misra2006}. \citet{misra2006} utilizes non-linear analysis of RXTE observations of GRS 1915's behaviour to categorize each of the classes into 3 main groups of behaviour: stochastic, deterministic and chaotic. Our network did not show equivalent separation with the exception of the stochastic grouping. Instead, we found that portions of each class were separated into different types of behaviour by the network. The grouping of classes by one type of behaviour by \citet{misra2006} seems too broad to apply to a whole class. We suggest instead that it would be more accurate to describe the proportion of time a class spends in these groupings as opposed to a class being constrained to only one of these groupings.

\subsection{Limitations} \label{limitations}

One limitation of the network built within this paper is that it often does deviate quite drastically from the original observation, even if it grasps the major important behaviours. Part of the cause of this is the limited size of the network. The choice to restrain the network's size to that of within this paper was taken due to the increasingly high computational cost of training the network as it expands in size. The network could be easily expanded to be composed of a far greater number of layers with the goal of more accurately recreating the observational data, especially if one were willing to invest more computation time into training.

Another flaw is that reconstructions tend to be inaccurate in the first and final 10 seconds of the segment; this is caused by the auto-encoder losing information on the edges when a pooling layer is applied. This manifests as the decoder reconstructs the data; it tends toward recreating the same shapes when up-sampling information due to insufficient information regarding the start and end of the segment. While this is unlikely to have had a major effect on the network's ability to find important behaviour, it does represent a limitation of the technique. Fig. \ref{fig:reconstruct} shows two examples of intensity light curve segments from the RXTE data set and the corresponding reconstructions from the auto-encoder which illustrates this effect. 

\subsection{Potential Improvements}

This technique could be further improved by the replacing of the auto-encoder with a similar concept called a transformer \citep{dosovitskiy2020image}. Transformers were developed by Google to analyze sequential data in natural language processing (NLP) and has been shown to outperform convolutional neural networks (similar to that used in this paper). This high level of performance is not purely limited to NLP. Transformers have already been used in an astronomical context for several different applications including: galaxy morphological classification \citep{lin2021galaxy}, photo-metric classification \citep{allam2021paying}, modelling physical systems \citep{geneva2022transformers}, and removing noise from time series \citep{morvan2022don}. The primary advantage of transformers over techniques like auto-encoders is the innovation of "self-attention". Self-attention enables transformers to take into account the position of features on multiple different time scales. This innovation would be particularly effective in distinguishing GRS 1915's behaviour which features similar behaviour on different time scales.

Another improvement that could be implemented is a new innovation in neural networks that mimics sleep in the brain \citep{Golden2022}. GRS1915 exhibits an enormous variety in behaviour with a heavy imbalance in the occurrence of a given type of behaviour. The reproduction of a particular type of behaviour can be broadly defined as a task that the network must perform. As the network is trained on a task (e.g. reproducing the peaks of the $\rho$ class), it learns how best to execute this task. However, when the network is trained to perform a different task, the network often performs considerably poorer when prompted to perform the original task. This is called catastrophic forgetting. The traditional way of mitigating this effect is by interleaving tasks during training so that the network is continually being retrained to perform tasks (which was the method undertaken in this paper). \citet{Golden2022} outlines that, by using a method mimicking sleep, this interleaving of tasks can be replaced by putting the network to "sleep" between training of tasks by reactivating neurons that are frequently used in a previous task. This technique has two advantages: 1) you do not need to store data from previous tasks to perform the sleep interleaving, and 2) new tasks can be more robustly learnt (neurons that aren't used frequently in task 1 are likely to be assigned to task 2). In the case of GRS1915, certain behaviour is hard for the network to reproduce as it appears very infrequently in the data. Use of sleep in the neural networking training would allow for iterative optimization for different characteristic patterns.

\section{Conclusions} \label{conclusions}
We created an auto-encoder machine learning network that is capable of organizing the behaviour of GRS 1915+105’s variability classes. It arrives at the same broad conclusions as B00, using the same information: time domain information in intensity and two color ratios. This affirms the system defined by B00 and provides a valuable confirmation of measurable differences between the classes.

We find that the majority of the classes meaningfully stand distinct from one another, even when they compromise a small amount of the data set. We find that there are reasonable grounds for considering the many variability patterns to be composed of 3 major types of behaviour: random movements, flaring and dips, and lengthening transitions between high and low intensities. These 3 types of behaviour are likely to be similar physical phenomena. We suggest that investigation of these shorter time periods of common behaviour could reveal similarities in the evolution of the physical system.

We recommend further investigations into similarities between $\delta$ and $\phi$  in particular as well as similarities between $\nu$ and $\alpha$. The physical cause behind similarities between classes may be revealed upon inspection of the power density spectra of the observations. Direct incorporation of power density spectra into the auto-encoder would be another valid path.

We also recommend the application of this technique to the system IGR J17091–3624. IGR J17091–3624 has historically exhibited similar behaviour to that of GRS 1915, albeit at a lower count rate \citep{Altamirano2011}. It recently had another outburst after a prolonged period of relative inactivity. This makes it an ideal candidate for observations as further study of GRS 1915's variability cannot be undertaken due to GRS 1915's novel obscured state \citep{balakrishnan2021,miller2020,neilsen2020}.

\color{black}
\section{Acknowledgements}
This research has made use of data and software provided by the High Energy Astrophysics Science Archive Research Center (HEASARC), which is a service of the Astrophysics Science Division at NASA/GSFC. Some of the network training in this paper was conducted on the Smithsonian High Performance Cluster (SI/HPC), Smithsonian Institution. B.J.R. and J.F.S. acknowledge support from  NASA grant no. GO9-20041X. D.H. is supported by the Women In Science Excel (WISE) programme of the Netherlands Organisation for Scientific Research (NWO). \myRed{We thank the anoynomous referee for their insightful comments.}

\section{Data Availability}
The data and code underlying this article are archived in Zenodo (\url{https://doi.org/10.5281/zenodo.7547328}). A streamlined variant program that can be run by readers to view the interactive UMAP projections and associated data can be found at \url{https://github.com/bjricketts/grs1915-auto-encoder.git}.


\bibliographystyle{mnras.bst}
\bibliography{biblio}


\appendix
\section{Reconstruction loss by class}
\begin{table}
\centering
\begin{tabular}{ |c|c|c|c| }
 \hline
Class & Intensity & HR1  & HR2  \\ 
 \hline
$\alpha$ & 3.35 & 0.73 & 0.74 \\
$\beta$ & 44.63 & 4.65 & 4.29 \\
$\chi$ & 3.76 & 0.80 & 0.70 \\
$\delta$ & 8.08 & 0.90 & 0.55 \\
$\gamma$ & 12.07 & 1.88 & 0.58 \\ 
$\lambda$ & 52.39 & 5.26 & 3.45 \\
$\kappa$ & 76.02 & 8.79 & 4.84 \\
$\mu$ & 59.40 & 8.35 & 2.08 \\
$\nu$ & 7.79 & 1.65 & 1.49 \\ 
$\omega$ & 58.35 & 7.00 & 3.69 \\
$\phi$ & 1.52 & 0.37 & 0.24 \\
$\rho$ & 54.83 & 5.95 & 3.92 \\
$\theta$ & 16.24 & 2.04 & 1.73 \\
  \hline

\end{tabular}
\caption{Table of the median reconstruction loss by class as defined by equation \ref{eq:wmse}. All reconstruction losses are shown only to two decimal places.}
\label{tab:reconstruction_errs}
\end{table} 

Table \ref{tab:reconstruction_errs} shows the average reconstruction loss of each class. Reconstruction losses of intensity are consistently higher due to considerably smaller errors in observation. Low sample size classes tend to be more poorly reconstructed. The network is unable to identify and recreate the features at the same level as more frequent classes because of this low sample size. While this is obviously not desirable, the network does manage to split these observations apart from one another despite this. There is a good chance that the architecture of the network may struggle specifically with reconstruction but manage to still capture features in the code, leading to this disparity between high errors and performance in relating segments together.

\section{Histogram plots of latent variables} \label{appendix:codes}

\begin{figure}
    \centering
    \includegraphics[width=\linewidth]{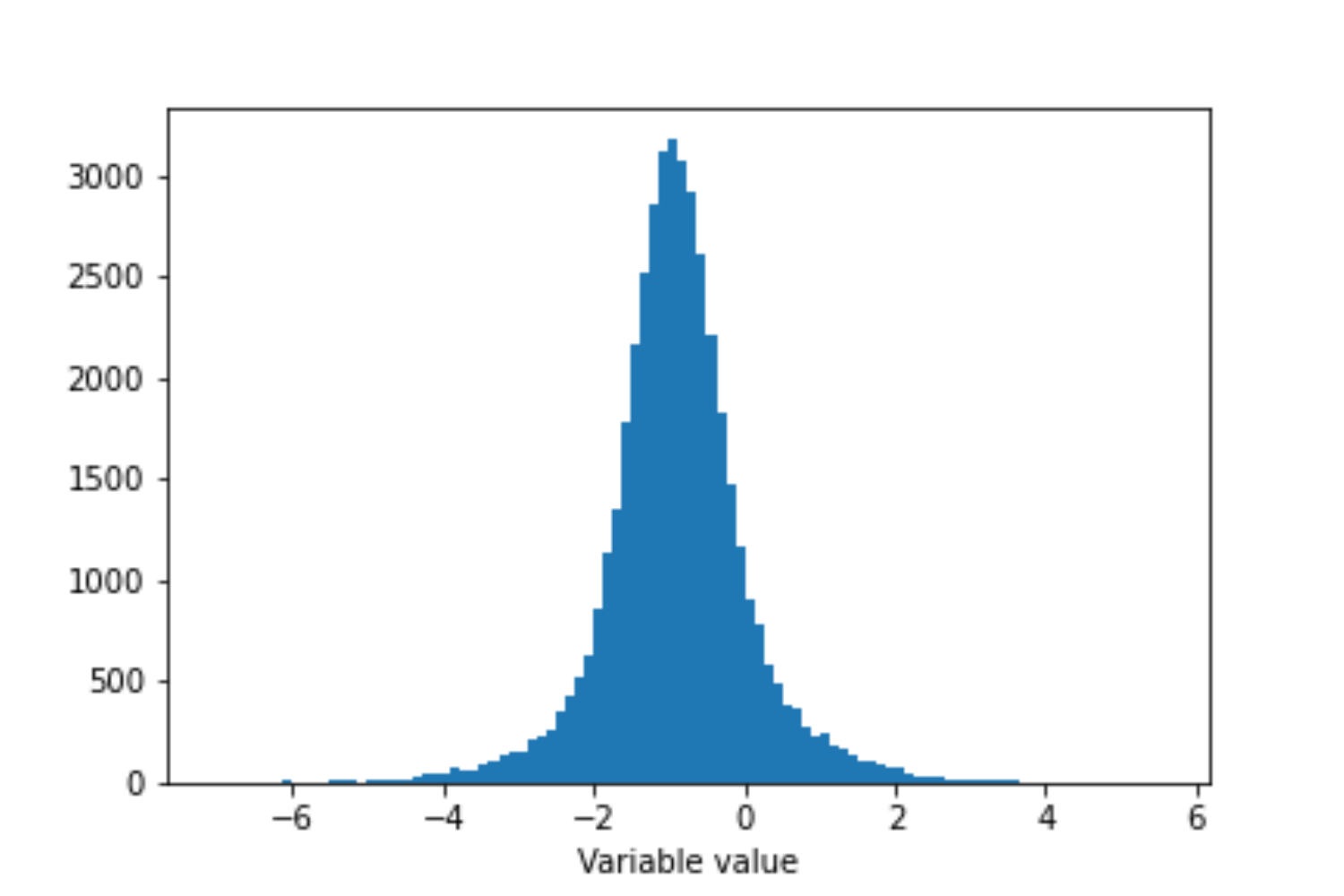}
    \caption{A histogram of one of the variables in the latent variables from intensity produced by the network. While the overall distribution is Gaussian, it should be noted that classes do not all follow this distribution.}
    \label{fig:hist_vars}
\end{figure}

\begin{figure*}
    \centering
    \includegraphics[width=\linewidth]{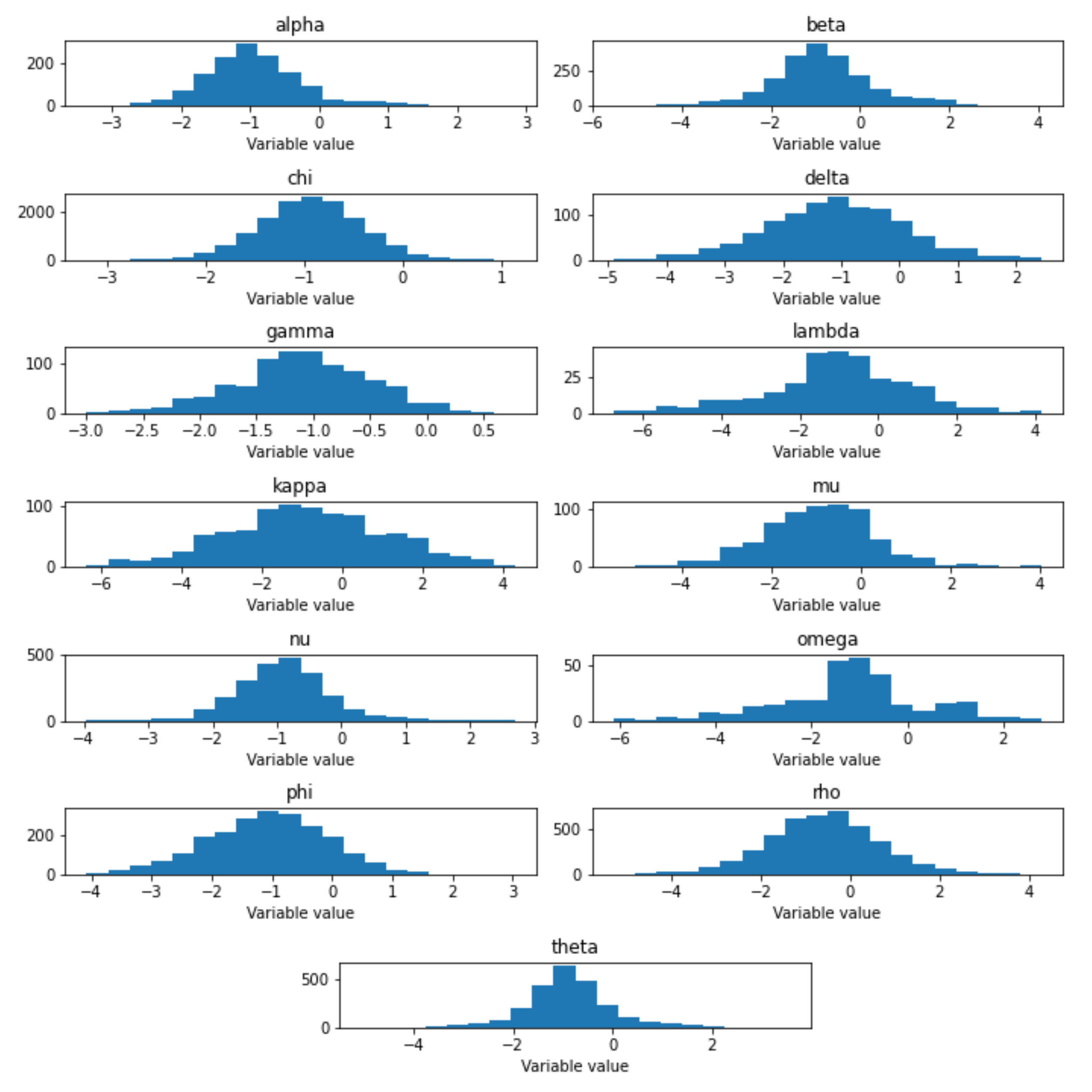}
    \caption{Histograms of each class' distribution in the same code variable as Fig. \ref{fig:hist_vars}.}
    \label{fig:hist_vars_all}
\end{figure*}

This appendix contains information about the variables inside the code to attempt to provide an insight into the "black box" of the network. The distribution of latent variables can be useful for assessing the network’s behaviour and the structure of the UMAP projection.

Fig. \ref{fig:hist_vars} shows an example distribution of one of the latent variables produced by the network. The latent variables used for this example were produced from intensity light curves. The aggregate distributions tend to be Gaussian, with a few exceptions. However, when a latent variable’s distribution is separated by class, clear skew emerges in the classwise profile. This demonstrates the network’s ability to segregate between classes meaningfully using informative latent variables.

Fig. \ref{fig:hist_vars_all} shows this bias in distributions for the same code variable as in Fig. \ref{fig:hist_vars}.
\label{lastpage}
\end{document}